\documentclass[useAMS,usenatbib]{mn2e}
\usepackage{natbib}
\usepackage{amsmath,amsfonts}
\usepackage{epsfig}
\usepackage{mathptmx}
\usepackage{longtable}

\def\reff@jnl#1{{\rm#1\/}}

\def\aj{\reff@jnl{AJ}}                  
\def\araa{\reff@jnl{ARA\&A}}            
\def\apj{\reff@jnl{ApJ}}                        
\def\apjl{\reff@jnl{ApJ}}               
\def\apjs{\reff@jnl{ApJS}}              
\def\ao{\reff@jnl{Appl.Optics}}         
\def\apss{\reff@jnl{Ap\&SS}}            
\def\aap{\reff@jnl{A\&A}}                       
\def\apjl{\reff@jnl{ApJ}}               
\def\aapr{\reff@jnl{A\&A~Rev.}}         
\def\aaps{\reff@jnl{A\&AS}}             
\def\azh{\reff@jnl{AZh}}                        
\def\baas{\reff@jnl{BAAS}}              
\def\jrasc{\reff@jnl{JRASC}}            
\def\memras{\reff@jnl{MmRAS}}           
\def\mnras{\reff@jnl{MNRAS}}            
\def\pra{\reff@jnl{Phys. Rev. A}}         
\def\prb{\reff@jnl{Phys. Rev. B}}         
\def\prc{\reff@jnl{Phys. Rev. C}}         
\def\prd{\reff@jnl{Phys. Rev. D}}         
\def\prl{\reff@jnl{Phys. Rev. Lett}}      
\def\pasp{\reff@jnl{PASP}}              
\def\pasj{\reff@jnl{PASJ}}              
\def\qjras{\reff@jnl{QJRAS}}            
\def\skytel{\reff@jnl{S\&T}}            
\def\solphys{\reff@jnl{Solar~Phys.}}    
\def\sovast{\reff@jnl{Soviet~Ast.}}     
\def\ssr{\reff@jnl{Space~Sci.Rev.}}     
\def\zap{\reff@jnl{ZAp}}                        
\def\nat{\reff@jnl{Nature}}             

\def\p#1by#2{{\partial{#1} \over \partial{#2}}}
\def\pp#1by#2#3{{\partial^2{#1} \over \partial{#2}\partial{#3}}}
\def\d#1by#2{{{\rm d}{#1} \over {\rm d}{#2}}}
\def\dd#1by#2#3{{{\rm d}^2{#1} \over {\rm d}{#2}{\rm d}{#3}}}

\title[AMI-LA observations of Serpens]{AMI-LA radio continuum observations of \emph{Spitzer} c2d small clouds and cores: Serpens region\thanks{We request that any reference to this paper cites ``AMI Consortium: Scaife et~al. 2011''.}}

\label{firstpage}

\author[Scaife et~al.]{
 AMI Consortium: Anna M. M. Scaife$^1$\thanks{email: ascaife@cp.dias.ie},
 Jennifer Hatchell$^2$, 
 Rachael E. Ainsworth$^1$,
\newauthor
 Jane V. Buckle$^{3,4}$,
 Matthew Davies$^3$,
 Thomas M. O. Franzen$^3$,
 Keith J. B. Grainge$^{3,4}$,
\newauthor
 Michael P. Hobson$^3$,
 Natasha Hurley-Walker$^3$,
 Anthony N. Lasenby$^{3,4}$,
 Malak Olamaie$^3$,
\newauthor
 Yvette C. Perrott$^3$,
 Guy G. Pooley$^3$,
 John S. Richer$^{3,4}$,
 Carmen Rodr{\'i}guez-Gonz{\'a}lvez$^3$,
\newauthor
 Richard D. E. Saunders$^{3,4}$,
 Michel P. Schammel$^3$,
 Paul F. Scott$^3$,
 Timothy Shimwell$^3$,
\newauthor
 David Titterington$^3$,
 Elizabeth Waldram$^3$.
\vspace{0.03in}\\
$^1$ Dublin Institute for Advanced Studies, 31 Fitzwilliam Place,
     Dublin 2, Ireland\\
$^2$ School of Physics, University of Exeter, Stocker Road, Exeter EX4 4QL\\
$^3$ Astrophysics Group, Cavendish Laboratory, J J Thomson Avenue,
     Cambridge CB3 0HE\\
$^4$ Kavli Institute for Cosmology, Cambridge, Madingley Road,
     Cambridge, CB3 0HA\\
}

\date{Accepted ---; received ---; in original form \today}

\pagerange{\pageref{firstpage}--\pageref{lastpage}}

\pubyear{2010}

\begin{document}
\maketitle

\begin{abstract}
We present deep radio continuum observations of the cores identified as deeply embedded young stellar objects in the Serpens molecular cloud by the \emph{Spitzer} c2d programme at a wavelength of 1.8\,cm with the Arcminute Microkelvin Imager Large Array (AMI-LA). These observations have a resolution of $\approx 30$\,arcsec and an average sensitivity of 19\,$\mu$Jy\,beam$^{-1}$. The targets are predominantly Class I sources, and we find the detection rate for Class~I objects in this sample to be low (18\%) compared to that of Class~0 objects (67\%), consistent with previous works. For detected objects we examine correlations of radio luminosity with bolometric luminosity and envelope mass and find that these data support correlations found by previous samples, but do not show any indiction of the evolutionary divide hinted at by similar data from the Perseus molecular cloud when comparing radio luminosity with envelope mass. We conclude that envelope mass provides a better indicator for radio luminosity than bolometric luminosity, based on the distribution of deviations from the two correlations. Combining these new data with archival 3.6\,cm flux densities we also examine the spectral indices of these objects and find an average spectral index of $\bar{\alpha}_{3.5}^{1.8}=0.53\pm1.14$, consistent with the canonical value for a partially optically thick spherical or collimated stellar wind. However, we caution that possible inter-epoch variability limits the usefulness of this value, and such variability is supported by our identification of a possible flare in the radio history of Serpens~SMM~1. 
\end{abstract}


\section{Introduction}
Recent radio follow-up of the \emph{Spitzer} catalogue of candidate low-luminosity embedded protostars has provided substantial new data with which to constrain the correlations observed between radio luminosity and the other physical parameters of protostellar objects in the low-luminosity limit (AMI Consortium: Scaife et~al. 2011a; 2011b, hereafter Papers I \& II). These data have also provided information on a handful of very low-luminosity objects (VeLLOs; Young et~al. 2004) which are of particular interest because their luminosities violate the lower limit set by steady accretion models (e.g. Shu 1977), sometimes by over an order of magnitude (Evans et~al. 2009). This discrepancy provides evidence for alternative models of protostellar evolution, such as non-steady accretion (Kenyon \& Hartmann 1995; Young \& Evans 2005; Enoch et~al. 2007). 

This paper will deal predominantly with low mass protostars in the earliest stages of their evolution. Class~0 objects represent the youngest protostars, generally distinguished from gravitationally bound pre-stellar cores by the presence of a detectable IR source, a compact radio counterpart, or a collimated molecular outflow (Andr{\'e} et~al. 1999). Such objects have yet to accrete significant mass from their surrounding envelopes so that $M_{\rm{env}}>M_{\ast}$. They are very faint at optical and Near Infra-Red wavelengths, but have significant sub-millimetre emission. Their SEDs are normally well characterized by single blackbodies, typically with $10\leq T_{\rm{d}} \leq 15$\,K (Hatchell et~al. 2007), and they possess energetic outflows driven by their accretion. Class~I objects are considered to be a slightly more evolved stage of protostellar evolution where the accretion rate has declined, and this is reflected in their less powerful outflows (Bontemps et~al. 1996) and a more clearly visible IR counterpart. In addition they have developed, or are in the process of developing, circumstellar disks. The presence of such disks alters the form of the SED, broadening it and flattening the Rayleigh-Jeans tail due to dust grain growth. Class~II objects (Classical T~Tauri stars) represent the start of the main sequence phase of stellar evolution, where infall is complete and the central object is embedded in an optically thick disc. Although protostellar evolution can be broadly encompassed by these definitions there is no strict division between the different stages, and classification can be complicated by a number of observational factors.

Unlike the correlations with bolometric and infra-red luminosity similar to those already known for higher luminosity ($>10^3$\,L$_{\odot}$; Anglada 1995) protostellar sources derived in Paper I, the observed correlation of radio luminosity with envelope mass observed in Paper II suggested an evolutionary distinction between Class 0 and Class I objects. This distinction is potentially explained by a difference in the origin of radio emission from protostellar objects between these classes, with Class 0 sources producing radio emission through an intrinsic mechanism but Class I objects being more heavily influenced by local environmental conditions. However the poor detection rate for Class I objects at 1.8\,cm prevents a conclusive statement about this difference to be made.

The Serpens molecular cloud is a highly extinguished, active near-by star formation region at a distance of $D=260\pm10$\,pc (Strai{\v z}ys et~al. 1996). The main Serpens cluster ($\simeq 6'$ extent), sometimes referred to as ``Cluster A'' (Harvey et~al. 2007), has been extensively studied in the infrared and submillimetre (see e.g. Davis et~al. 1999; Casali et~al. 1993; Enoch et~al. 2007; Graves et~al. 2010) although there are few surveys of the full extent of the optical extinction attributed to the larger molecular cloud of 10\,deg$^2$ (Cambr{\'e}sy et~al. 1999). The region of active star formation approximately $45'$ to the south of the Serpens core is known as Serpens/G3-G6 (Cohen \& Kuhi 1979) or ``Cluster B'' (Enoch et~al. 2007; Harvey et~al. 2006). Both the Serpens core and the G3-G6 region have extensive multi-frequency observations available, including radio surveys at longer wavelengths (Eiroa et~al. 2005, Serpens core; Djupvik et~al. 2006, Serpens/G3-G6). The Serpens cloud contains a large proportion of Class I and II young stellar objects (YSOs) compared with the Perseus molecular cloud, considered in Paper II, although these are found predominantly not in the clusters but in the extended cloud (Harvey et~al. 2007), supporting the idea that the clusters, ``A'' and ``B'', are younger than the surrounding cloud.

In this paper we extend our radio follow-up of the \emph{Spitzer} catalogue of low-luminosity embedded protostars (Dunham et~al. 2008, hereafter DCE08) with the Arcminute Microkelvin Imager Large Array (AMI-LA; Papers I \& II) to cover the Serpens region. In \S~\ref{sec:sample} we describe the sample; in \S~\ref{sec:obs} we briefly introduce the telescope and describe the data reduction procedures utilised; in \S~\ref{sec:res} we describe the results of the observations towards the Cluster A, Cluster B and Serpens Filament individually and review the complementary data at other wavelengths. In \S~\ref{sec:corr} we draw correlations between the characteristic parameters of these objects, and in \S~\ref{sec:disc} we discuss the implications of these results with reference to detection statistics for differing protostellar classes, radio spectral indices and, where measured, outflow momentum fluxes. In \S~\ref{sec:conc} we draw our conclusions.

\begin{figure}
\includegraphics[width=0.5\textwidth]{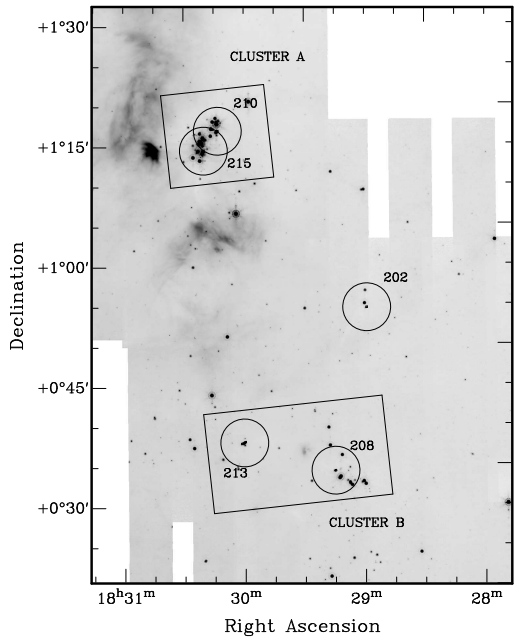}
\caption{\emph{Spitzer} 24\,$\mu$m map of Serpens with the FWHM of the primary beam for the AMI-LA fields, listed in Table~\ref{tab:obs}, shown as circles. The locations of the regions known as Cluster A and Cluster B are indicated. \label{fig:serpens}}
\end{figure}

\section{The Sample}
\label{sec:sample}
In this work we observe a sample of candidate deeply embedded protostellar cores from the \emph{Spitzer} catalogue of DCE08. Candidates for embedded objects from this catalogue were ranked by DCE08 as belonging to one of six ``Groups'', with Group~1 being those most likely to be true embedded objects, and Group~6 those least likely. The sample of objects observed here comprises all objects classified as Groups 1--3 in the Serpens region. Group 1 objects are sources which have been confirmed as embedded protostars independently of their \emph{Spitzer} identification, Group 2 objects are not confirmed as embedded in a dense core but have ancillary data confirming a molecular outflow, and Group 3 are not confirmed as embedded and do not have outflow data. We identify the candidates from this catalogue by their catalogue number, i.e. DCE08-$nnn$, in column [1] of Table~\ref{tab:sample}. The coordinates of the individual candidates are listed in columns [2] and [3] of Table~\ref{tab:sample}, along with their Group and infrared luminosity from Dunham et~al. (2008) in columns [4] and [5]. Three sources are found in the Cluster A region (DCE08-210, -211 \& -215) and four in the Cluster B region (DCE08-208 \& 212--214). The final source, DCE08-202, is outside the cluster regions but is found within the elongated sub-mm filament identified in Enoch et~al. (2007) to the north of Cluster B, referred to as the Serpens filament. The locations of the sources listed in Table~\ref{tab:sample} are shown in Fig.~\ref{fig:serpens}.

The source list in Table~\ref{tab:sample} has been cross-referenced with the larger \emph{Spitzer} cores to disks catalogue of Evans et~al. 2009 (hereinafter EDJ09) in order to provide bolometric luminosities and temperatures, and 3.6\,$\mu$m flux densities, as well as the Serpens \emph{Spitzer} catalogue of Harvey et~al. (2007; hereafter HMH07). 

Hatchell et~al. (2007) used three criteria to distinguish Class~I from Class~0: firstly that the bolometric temperature, $T_{\rm{bol}}>70$\,K; secondly that the ratio of bolometric to sub-mm luminosity, $L_{\rm{bol}}/L_{\rm{smm}}>3000$; and thirdly that $S_{3.6\,\mu \rm{m}}/S_{850\,\mu \rm{m}}>0.003$. The final classification for a source was based on the majority result from these three criteria. Although no measure of the sub-mm luminosity is available for these sources from the literature, all of the objects in this sample would be classified as Class I by both of the two alternative indicators, giving a majority without the need for a third indicator. These indicators have been calculated using the physical data from Evans et~al. (2009).  
However Enoch et~al. (2009) classify three of these objects as Class~0 and it is likely that they are borderline cases as the distinction between Class~0 and Class~I is not exact. The classifications from Enoch et~al. (2009) are listed in Column [11] of Table~\ref{tab:sample}.

From the correlation of radio luminosity and bolometric luminosity derived in Paper II we can predict the expected 16\,GHz radio flux densities of these sources from the bolometric luminosities in Evans et~al. (2009). These predictions are shown in Column [13] of Table~\ref{tab:sample} and have uncertainties of a factor of $\approx2$. However, although this general trend is observed for \emph{detected} sources, we note that a high proportion of Class I objects are un-detected in the radio.

\begin{table*}
\caption{The AMI-LA sample of embedded protostellar sources selected from the catalogue of Dunham et~al. (2008). Columns are [1] source number from the Dunham et~al. (2008) catalogue; [2] Right Ascension of source in J2000 coordinates; [3] Declination of source in J2000 coordinates; [4] candidate Group from Dunham et~al. (2008); [5] IR luminosity from Dunham et~al. (2008); [6] identifier from Evans et~al. (2009); [7] bolometric luminosity from Evans et~al. (2009); [8] bolometric temperature from Evans et~al. (2009); [9] identification from Harvey et~al. (2007); [10] identifier from Enoch et~al. (2009); [11] classification from Enoch et~al. (2009); [12] envelope mass from Enoch et~al. (2009); [13] predicted 1.8\,cm flux density based on derived correlation between bolometric luminosity and radio luminosity (Paper~II). \label{tab:sample}}
\begin{tabular}{cccccccccccccc}
\hline\hline
[DCE08] & RA & Dec. & Group & $L_{\rm{IR}}$ &  [EDJ09] & $L_{\rm{bol}}$ & $T_{\rm{bol}}$ & [HMH07] & Bolo & Class & $M_{\rm{env}}$ & $S_{1.8\,\rm{cm}}^{\rm{pred}}$ \\
 & (J2000) & (J2000) & & (L$_{\odot}$) &  & (L$_{\odot}$) & (K) & & & & (M$_{\odot}$) & (mJy)  \\
\hline
202& 18 28 44.78 &+00 51 25.9  &3  &   0.061	&527 &0.078	&180	&24 & 3  & I &$0.24\pm0.02$ & 0.069 \\
208& 18 29 09.05 &+00 31 27.8  &1  &   0.009	&577 &0.024     &410	&75 & 15 & 0 &$1.16\pm0.02$ & 0.040 \\
210& 18 29 49.63 &+01 15 22.0  &2  &   0.347	&637 &2.1       &97	&141& 23 & 0 &$7.98\pm0.07$ & 0.317 \\
211& 18 29 51.98 &+01 15 38.2  &3  &   0.036	&-   &4.0       &110	&-  & -  & I$^a$ & - & 0.426\\
212& 18 29 52.51 &+00 36 11.9  &3  &   0.015    &650 &0.85      &100	&154& 24 & 0 &$0.54\pm0.03$ & 0.209 \\
213& 18 29 53.04 &+00 36 06.8  &2  &   0.185    &653 &0.57      &860	&157& -  & I$^a$ & - & 0.173\\
214& 18 29 54.31 &+00 36 01.4  &2  &   0.105    &661 &0.64      &110	&166&24  & I &$0.48\pm0.05$ & 0.182 \\
215& 18 29 57.67 &+01 13 04.4  &2  &   0.018 	&676 &0.55      &820	&181&25/28&I & - & 0.170\\
\hline
\end{tabular}
\begin{minipage}{\textwidth}{
$^a$ Classification based on the criteria of Hatchell et~al. (2007).
}
\end{minipage}
\end{table*}

\section{Observations}
\label{sec:obs}

AMI comprises two synthesis arrays, one of ten 3.7\,m
antennas (SA) and one of eight 13\,m antennas (LA),
both sited at the Mullard Radio Astronomy Observatory at Lord's Bridge, Cambridge (AMI Consortium: Zwart
et~al. 2008). The telescope observes in 
the band 13.5--17.9\,GHz with eight 0.75\,GHz bandwidth channels. In practice, the two
lowest frequency channels (1 \& 2) are not generally used due to a lower response in this frequency range and interference from geostationary
satellites. The data in this paper were taken with the AMI Large Array (AMI-LA).

Observations of the 8 objects listed in Table~\ref{tab:sample} were made with the AMI-LA between December 2010 and January 2011. The eight targets were observed in five separate pointings:DCE08-202, DCE08-208 and DCE08-215 were considered sufficiently separated to be observed as separate pointings; DCE08-210 and DCE08-211 are separated by $\simeq 40''$ and were therefore observed as a single pointing; DCE08-212, 213 \& 214 have a maximum separation of just over 1\,arcmin and were therefore also observed as a single pointing. These objects lie within the declination range $0-2^{\circ}$, outside the standard observing range of the AMI telescope; below $\delta\simeq 15^{\circ}$ significant interference from geostationary satellites is experienced. The effect of violating this declination limit is a higher noise level than the telescope specifications due to both the heavier data flagging required to remove satellite interference and the residual contamination.
\begin{table}
\caption{AMI-LA frequency channels and primary calibrator flux densities measured in Jy.\label{tab:cals}}
\begin{tabular}{lcccccc}
\hline \hline
Channel No. & 3 & 4 & 5 & 6 & 7 & 8\\
Freq. [GHz] & 13.88 & 14.63 & 15.38 & 16.13 & 16.88 & 17.63 \\
\hline
3C48  & 1.85 & 1.75 & 1.66 & 1.58 & 1.50 & 1.43 \\
3C286 & 3.60 & 3.54 & 3.42 & 3.31 & 3.21 & 3.11 \\
3C147 & 2.75 & 2.62 & 2.50 & 2.40 & 2.30 & 2.20 \\
\hline
\end{tabular}
\end{table}
\begin{table}
\caption{AMI-LA observations. Columns are [1] source number from the Dunham et~al. (2008) catalogue; [2] AMI-LA flux calibrator; [3] AMI-LA phase calibrator; [4] rms noise measured from recovered map; [5] major axis of AMI-LA synthesized beam; [6] minor axis of AMI-LA synthesized beam.  \label{tab:obs}}
\begin{tabular}{lccccc}
\hline\hline
[DCE08]$^{\dagger}$ & $1^{\circ}$ & $2^{\circ}$ & $\sigma_{\rm{rms}}$ & $\Delta \theta_{\rm{max}}$ & $\Delta \theta_{\rm{min}}$ \\
&&&($\frac{\mu \rm{Jy}}{\rm{beam}}$) & (arcsec) & (arcsec) \\
\hline
$\begin{array}{c}
{\textbf{202}}
\end{array}$ 
& 3C286& J0329+2756 & 19  & 41.4 & 28.2 \\
$\begin{array}{c}
{\textbf{208}}
\end{array}$ 
& 3C286& J0329+2756 & 16  & 52.6 & 26.6 \\
$\left.\begin{array}{c}
{\textbf{210}}\\
211
\end{array} \right\}$
& 3C48 & J0329+2756 & 19  & 48.2 & 26.3 \\
$ \left.\begin{array}{c}
212\\
{\textbf{213}}\\
214
\end{array} \right \}$
& 3C286& J0329+2756 & 20  & 40.2 & 29.8 \\
$\begin{array}{c}
{\textbf{215}}
\end{array}$ 
& 3C48 & J0329+2756 & 22  & 48.5 & 25.9 \\
\hline
\end{tabular}
\begin{minipage}{0.5\textwidth}{
$^{\dagger}$ Numbers in bold indicate field designations as shown in Fig.~\ref{fig:serpens}
}
\end{minipage}
\end{table}

AMI-LA data reduction is performed using the local software tool \textsc{reduce}. This applies
both automatic and manual flags for interference, 
shadowing and hardware errors, Fourier transforms the correlator data to synthesize frequency
channels and performs phase and amplitude
calibrations before output to disc in \emph{uv} FITS format suitable for imaging in
\textsc{aips}\footnote{\tt http://www.aips.nrao.edu/}. Flux (primary) calibration is performed using short observations of 3C286, 3C48 and 3C147. We assume I+Q flux densities for these sources in the
AMI-LA channels consistent with the updated VLA calibration scale (Rick Perley, private comm.), see Table~\ref{tab:cals}. Since the AMI-LA measures
I+Q, these flux densities 
include corrections for the polarization of the calibrator sources. A correction is
also made for the changing air mass over the observation. From
other measurements, we find the flux calibration is accurate to better than
5 per cent (AMI Consortium: Scaife et~al. 2008; AMI Consortium:
Hurley--Walker et~al. 2009). Additional phase (secondary) calibration is done using interleaved observations of
calibrators 
selected from the Jodrell Bank VLA Survey (JVAS; Patnaik et~al. 1992). After calibration, the phase is generally stable to
$5^{\circ}$ for channels 4--7, and
$10^{\circ}$ for channels 3 and 8. The FWHM of the primary beam of the AMI-LA is $\approx 6$\,arcmin at 16\,GHz. Due to their superior phase stability only channels $4-7$ were used for this work, resulting in an effective total bandwidth of 3\,GHz.

Reduced data were imaged using the AIPS data package. {\sc{clean}}
deconvolution was performed using the task 
{\sc{imagr}} which applies a differential primary beam correction to
the individual frequency channels to produce the combined frequency
image. Due to the low declination of this sample, uniform visibility weighting was used to improve the AMI-LA PSF. The AMI-LA is sensitive to angular scales from $\approx0.5-6$\,arcmin, although this varies as a function of declination, hour angle coverage and data flagging.

In what follows we use the convention: $S_{\nu}\propto \nu^{\alpha}$, where $S_{\nu}$ is
flux density (rather than flux, $F_{\nu}=\nu S_{\nu}$), $\nu$ is frequency and $\alpha$ is the spectral index. All errors quoted are 1\,$\sigma$. 

Where a known object is not detected by the AMI-LA an upper limit on the 1.8\,cm radio flux density of that source is given. These limits take into account the position of the source within the AMI-LA primary beam. Since source detection is performed in the maps before correction for the  primary beam this means that the true limit on the unattenuated flux density will be given by $S_{\rm{lim}}<5\,\sigma_{\rm{th}}/A(r)$, where $A(r)$ is the primary beam attenuation at a radial distance $r$ from the pointing centre and $\sigma_{\rm{th}}$ is the r.m.s. noise on the map. 
\begin{table*}
\caption{AMI-LA observations of the Serpens Core (Cloud A). Columns are [1] AMI-LA source number; [2] Right Ascension of radio source peak; [3] Declination of radio source
 peak; [4] Integrated 1.8\,cm flux density; [5] VLA association; [6] Integrated 
3.5\,cm flux density; [7] Spectral index from 3.5 to 1.8\,cm; [8] Associations from HMH07; [9] Bolometric luminosity from Enoch et~al. (2009a), values in italics from Evans et~al. (2009); [10] envelope mass from Enoch et~al. (2009a); \& [11] protostellar classification\label{tab:serpa}} 
\begin{tabular}{ccccccccccc}
\hline\hline 
AMI&RA&Dec&$S_{\rm{1.8\,cm,int}}$&VLA&$S_{\rm{3.5\,cm,int}}$&$\alpha_{3.5}^{1.8}$&[HMH07]&$L_{\rm{bol}}$&$M_{\rm{env}}$ & Cl.  \\ 
&(J2000) &(J2000) &(mJy) & &(mJy) & & &(L$_{\odot}$) &(M$_{\odot}$) & \\
\hline 
\multicolumn{11}{l}{\emph{DCE08-210}:}\\
1 &18 29 49.63  &+01 15 21.9 & $4.736\pm0.237$ & 7 & 7.54 & $-0.73\pm0.16$ &141   & 17.3$^{\dagger}$ & $7.98\pm0.07$ & 0\\

2 &$\left.\begin{array}{c}
18~29~51.06 \\
18~29~52.19 
\end{array} \right.$ &$\left.\begin{array}{c}
+01~15~33.9\\
+01~15~47.8
\end{array}  \right.$ & $1.301\pm0.067$ &$\left.\begin{array}{c}
10\\
8
\end{array} \right.$ & $\left.\begin{array}{c}
0.14\\
0.64
\end{array} \right.$& $0.81\pm0.10$ & $\left.\begin{array}{c}
150\\
-
\end{array} \right.$ & $\left.\begin{array}{c}
2.6\\
-
\end{array} \right.$ & $\left.\begin{array}{c}
1.23\pm0.12\\
-
\end{array} \right.$ & $\left.\begin{array}{c}
{\rm{I}}\\
-
\end{array} \right.$ \\
 - &18 29 52.85  &+01 14 56.0 & $<0.101$  & 11& 0.09 & $<0.18$ &155   & 1.2 & $0.74\pm0.07$ & I\\
 - &18 29 49.13  &+01 16 19.8 & $<0.102$  & - & - & - &137   & 4.4 & $1.53\pm0.15$ & I \\
 - &18 29 51.14  &+01 16 40.6 & $<0.110$  & 9 & 0.15 & $<-0.49$ &146   & 1.7 & $1.84\pm0.18$ & I\\
 3 &18 29 48.10  &+01 16 44.9 & $0.355\pm0.026$ & 5 & 0.28 & $0.38\pm0.44$ &135   & 2.1 & $3.72\pm0.37$ & 0\\
 4 &18 29 49.60  &+01 17 07.0 & $0.245\pm0.023$ & - & -    & - &139  &\emph{1.1} & -  & I$^{b}$\\
 - &18 29 56.87  &+01 14 46.5 & $<0.125$ & - & - & - &176   & 5.8 & $0.69\pm0.07$ & I\\
 - &18 29 57.72  &+01 14 05.7 & $<0.148$ & 16& 0.17 & $<-0.22$ &182   & 13.8 & $1.90\pm0.19$ & I\\
 - &18 29 58.77  &+01 14 26.2 & $<0.151$ & - & - & - &190   & 0.62& $2.00\pm0.20$ & I\\
 5$^{a}$&18 29 34.63  &+01 15 07.0 & $3.139\pm0.158$ & 2 & 2.17 & $0.58\pm0.05$ & - & - & - & -\\
 6$^{a}$&18 29 44.05  &+01 19 22.4 & $2.127\pm0.108$ & 3 & 2.10 & $0.02\pm0.005$ & - & - & - & -\\
\hline
\multicolumn{11}{l}{\emph{DCE08-215}:}\\
 7 &18 29 56.33  &+01 13 19.4 & $0.450\pm0.031$ & 13 & 0.12 & $2.09\pm0.00$ & 175 & 2.6 & $2.35\pm0.24$ & I\\
 - &18 29 57.67  &+01 13 04.4 & - & - & - & - &181  & 2.0 & $2.56\pm0.03$ & I\\
 - &18 29 57.89  &+01 12 46.2 & - &17'& 0.18 & - &187   & \emph{32.0} & - & - \\
 8 &18 29 57.84  &+01 12 37.8 & $1.000\pm0.050$ & 17& 1.82 & $-1.05\pm0.02$ &186   & \emph{6.1} & - & -\\
 - &18 29 59.92  &+01 13 11.6 & - & 19& 0.11 & - &198   & \emph{4.0} & - &- \\
 - &18 30 00.70  &+01 13 01.4 & - & 15& 0.10 & - &203  & 1.2 & $2.56\pm0.03$ & 0\\
 - &18 29 57.72  &+01 14 05.7 & $<0.118$ & 16& 0.17 & - &182   & 13.8 & $1.90\pm0.19$ & I\\
 - &18 29 59.56  &+01 11 59.0 & $<0.122$ & - & - & - &197   & 1.8 & $1.14\pm0.11$ & I\\
 - &18 29 58.77  &+01 14 26.2 & $<0.128$ & - & - & - &190   & 0.62 & $2.00\pm0.20$ & I\\
 - &18 30 02.73  &+01 12 28.2 & $<0.128$ & - & - & - &208   & 2.2 & $1.32\pm0.13$ & I\\
 - &18 29 56.87  &+01 14 46.5 & $<0.138$ & - & - & - &176   & 5.8 & $0.69\pm0.07$ & I\\
 - &18 29 52.85  &+01 14 56.0 & $<0.162$ & 11 & 0.09 & - &155   & 1.2 & $0.74\pm0.07$ & I\\
\hline
\end{tabular}
\begin{minipage}{\textwidth}
$^{a}$ outside AMI-LA primary beam FWHM.\\
$^{b}$ classification from Kaas et~al. (2004) \\
$\dagger$ Enoch et~al. 2009b.
\end{minipage}
\end{table*}

\begin{table*}
\caption{AMI-LA observations of the Serpens Cloud B. Columns are [1] AMI-LA source number; [2] Right Ascension of radio source peak; [3] Declination of radio source
 peak; [4] Integrated 1.8\,cm flux density; [5] VLA association; [6] Integrated 
3.5\,cm flux density; [7] Spectral index from 3.5 to 1.8\,cm; [8] Associations from HMH07; [9] Bolometric luminosity from Enoch et~al. (2009a), values in italics from Evans et~al. (2009); [10] envelope mass from Enoch et~al. (2009a); \& [11] protostellar classification\label{tab:clb-obs}} 
\begin{tabular}{ccccccccccc}
\hline\hline
AMI & RA & Dec & $S_{\rm{1.8\,cm, int}}$ & VLA & $S_{\rm{3.5\,cm,int}}$ & $\alpha_{3
.5}^{1.8}$ & [HMH07] & $L_{\rm{bol}}$& $M_{\rm{env}}$ & Cl. \\
& (J2000) & (J2000) & (mJy) & & (mJy) & & & (L$_{\odot}$) & (M$_{\odot}$) & \\
\hline
\multicolumn{11}{l}{\emph{DCE08-208}:}\\
9 & 18 29 00.8 & +00 32 19 & $0.141\pm0.023$ & - & - & - & - & - & - & -\\
10 & 18 29 01.7 & +00 29 48 & $3.583\pm0.062$ & 4 & 3.30 & $0.13\pm0.04$  & 59 & 4.2$^{a}$ & - & II/III\\
11 & 18 29 06.4 & +00 30 38 & $1.204\pm0.121$ & 6 & 0.32 & $2.07\pm0.40$ & 68 & 6.5$^{a}$ & $2.77\pm0.14$ & 0\\
12 & 18 29 09.1 & +00 31 33 & $0.181\pm0.024$ & - & - & - & 75-A,B,C & 1.7 & $1.16\pm0.02$ & 0\\
- & 18 29 02.1 & +00 31 21 & $<0.0703$ & - & - & - & 60  & 0.09 & $0.17\pm0.02$ & I\\
- & 18 29 09.1 & +00 31 33 & $<0.0904$ & - & - & - & 63  & 0.12 & $0.22\pm0.02$ & I\\ 
\hline
\multicolumn{10}{l}{\emph{DCE08-213}:}\\
13 &18 29 52.2   &+00 38 51.5 & $0.589\pm0.067$ & - & - & - & - & - & - & -\\
14 &18 29 53.3   &+00 34 09.4 & $0.267\pm0.082$  & - & - & - & - & - & -& -\\
 - &18 29 53.04  &+00 36 06.8 & $<0.0950$ & - & - & - &157       & \emph{0.57} & - & -\\
 - &18 29 52.52  &+00 36 11.7 & $<0.0950$ & - & - & - &154   & 1.0 & $0.54\pm0.03$ & 0\\
 - &18 29 54.30  &+00 36 01.3 & $<0.0960$ & - & - & - &166   & 0.17 & $0.48\pm0.05$ & I \\
\hline
\end{tabular}
\begin{minipage}{\textwidth}
$^{a}$ from Harvey \& Dunham (2009).
\end{minipage}
\end{table*}

\begin{table*}
\caption{AMI-LA observations of the Serpens Filament. Columns are [1] AMI-LA source number; [2] Right Ascension of radio source peak; [3] Declination of radio source
 peak; [4] Integrated 1.8\,cm flux density; [5] VLA association; [6] Integrated 
3.5\,cm flux density; [7] Spectral index from 3.5 to 1.8\,cm; [8] Associations from HMH07; [9] Bolometric luminosity from Enoch et~al. (2009a), values in italics from Evans et~al. (2009); [10] envelope mass from Enoch et~al. (2009a); \& [11] protostellar classification\label{tab:filament}} 
\begin{tabular}{ccccccccccc}
\hline\hline
AMI & RA & Dec & $S_{\rm{1.8\,cm, int}}$ & VLA & $S_{\rm{3.5\,cm,int}}$ & $\alpha_{3
.5}^{1.8}$ & [HMH07] & $L_{\rm{bol}}$& $M_{\rm{env}}$ & Cl.\\
& (J2000) & (J2000) & (mJy) & & (mJy) & & & (L$_{\odot}$) & (M$_{\odot}$) & \\
\hline
\multicolumn{11}{l}{\emph{DCE08-202}:}\\
 - &18 28 44.8  &+00 51 26 & $<0.095$ & - & - & - & 24 & 0.04 & $0.24\pm0.02$ & I\\
 - &18 28 45.0  &+00 52 04 & $<0.098$ & - & - & - & 26 & 1.09 & $0.29\pm0.03$ & I\\
 - &18 28 44.0  &+00 53 38 & $<0.138$ & - & - & - & 23 & 0.19 & $0.26\pm0.03$ & I\\
\hline
\end{tabular}
\end{table*}

\section{Results}
\label{sec:res}

In this section we address each of the regions of the Serpens molecular cloud separately and examine the sources detected in the radio data individually. These examinations are made with reference to several archival data sets. Notably these include the Northern VLA Sky Survey (NVSS: Condon et~al. 1998) at 1.4\,GHz in the radio regime, the sub-mm catalogue of Enoch et~al. (2009) from the Bolocam instrument at 1.1\,mm and the Mid Infra-Red \emph{Spitzer} catalogue of Evans et~al. (2009), as well as the Serpens specific \emph{Spitzer} catalogue of Harvey et~al. (2007). 

Source type is characterized by radio spectral index with extra-galactic sources expected to have steeply falling spectra with frequency, $\alpha\geq -0.5$, and thermal sources expected to have shallow or rising spectra $-0.1\leq \alpha \leq 2$. Thermal sources are assumed to represent protostellar objects, where $\alpha=-0.1$ indicates optically thin free-free radio emission and $\alpha=2.0$ indicates optically thick emission. Values intermediate to these limits are assumed to be sources with partially optically thick media (see e.g. Reynolds 1986; Panagia \& Felli 1975). 

These details are consistent with data presented in Papers~I~\&~II where correlations were seen between the radio luminosity from objects identified as protostellar with a number of physical characteristics determined from other wavelength regimes. Particularly relevant to this work are the correlations with envelope mass and bolometric luminosity (Paper II) which allow us to make predictions of radio luminosity based on emission in other regimes,
\begin{eqnarray}
\nonumber \log[L_{\rm{1.8\,cm}} (\rm{mJy\,kpc}^2)] & = & -(1.74\pm0.18)\\
&&+(0.51\pm0.26)\log[L_{\rm{bol}} (\rm{L}_{\odot})],
\end{eqnarray}
and
\begin{eqnarray}
\nonumber\log[L_{\rm{1.8\,cm}} (\rm{mJy\,kpc}^2)] & = & -(2.23\pm0.65)\\
&&+(0.68\pm0.62)\log[M_{\rm{env}} (\rm{M}_{\odot})].
\end{eqnarray}
The predicted radio flux densities for the target sample presented here, based on bolometric luminosity, are listed in Table~\ref{tab:sample}.

\subsection{Serpens Core (Cluster A) Region}

Candidate cores DCE08-210, DCE08-211 and DCE08-215 are found in the region of the Serpens core. DCE08-210 is associated with the sub-mm core Serpens~SMM~1 and is closely separated from DCE08-211. At high resolution Serpens~SMM~1 can be resolved into a triple radio source, with components coincident with its core and outflows (Rodr{\'i}guez et~al. 1989; Curiel et~al. 1993) and is the most luminous object in this molecular cloud. DCE08-211 is associated with Serpens~SMM~10, which is an order of magnitude smaller in mass than its neighbour. DCE08-215 is located to the south-east of the previous two objects and is associated with Serpens~SMM~4, also known as EC95 (Eiroa \& Casali 1992) and thought to be the X-ray source Ser-X3 (Smith, G{\"u}del \& Benz 1999).

\subsubsection{Existing radio observations of cluster A}

The Serpens Cluster A region was previously surveyed at 3.5\,cm with the VLA by Eiroa et~al. (1993). Following the \emph{Spitzer} survey of this region it is now possible to match each of the 3.5\,cm detections with individual protostellar sources. These identifications are listed in Table~\ref{tab:serpa}. The flux densities have been adjusted to bring them on to the same flux scale as the AMI-LA, which follows the updated VLA calibration (Rick Perley private comm.). We identify radio detections at 8.5\,GHz from Eiroa et~al. (1993) with protostellar sources from Harvey et~al. (2007) when they are coincident to within a radius of $10''$. The protostellar objects detected at 3.5\,cm are in general deeply embedded objects, with no coincidence with known Class~II/III sources. Within the VLA fields there is only one source classified as embedded by Harvey et~al. (2007b) that is not detected in this sample, HMH07-195. Of the detected sources, two are not classified as embedded because they possess a 2MASS counterpart: HMH07-146 \& -182 but are identified as Class I by Enoch et~al. (2009); two are unclassified, most likely because both lack detections at 70\,$\mu$m: HMH07-182 \& -186; HMH07-198 is classified as cold and embedded by Harvey et~al. (2007b) however, it is not identified by Enoch et~al. (2009) as being part of a sub-mm core. Those sources which are not identified in the sub-mm are not necessarily more evolved objects but could instead be examples of deeply embedded sources with envelope masses below the detection threshold of current sub-mm surveys.

\subsubsection{DCE08-210}

The radio source detected here towards DCE08-210 is coincident with IRAS~18273+0113, also known as Serpens FIRS1 and Serpens~SMM~1. In the mid-infrared this object can be seen to encompass at least five separate sources. DCE08-210 is coincident with the near-infared source EC41 (Eiroa \& Casali 1992), which is also the ISOCAM source K258a (Kaas et~al. 2004). This object is thought to be a Class I source based on its $\alpha_{\rm{IR}}^{2-14}$ index and this classification agrees with the bolometric luminosity and temperature found by \emph{Spitzer} and given in Evans et~al. (2009), where it is designated [EJD2009]-637. However, using combined data from 2MASS, \emph{Spitzer} and Bolocam, Enoch et~al. (2009; 2009b) classify FIRS1 as a Class 0 YSO with a bolometric temperature of $56\pm12$\,K, a bolometric luminosity of $11\pm6$\,L$_{\odot}$ and an envelope mass of $7.98\pm0.07$\,M$_{\odot}$, consistent with its original IRAS classification (Hurt \& Barsony 1996). This is also supported by its observed molecular outflows, which are highly extended spatially and energetic (Davis et~al. 1999; Graves et~al. 2010), both properties more normally associated with Class~0 objects.

The closely associated object K258b, $13''$ to the south-west of K258a, and detected in the survey of Serpens by Harvey et~al. (2007) in four \emph{Spitzer} bands has been suggested to be simply a knot of extended emission (Kaas et~al. 2004).

The dust temperatures found from far-infrared and sub-mm studies of DCE08-210 (Hurt \& Barsony 1996; Davis et~al. 1999) are $30-40$\,K towards FIRS1. This is high for Class~0 objects, which typically have $10<T_{\rm{d}}<20$\,K (see e.g. Hatchell et~al. 2007; Visser et~al. 2002). It is uncertain therefore whether Serpens FIRS1, and therefore DCE08-210, is a late stage Class 0 YSO or an early stage Class I YSO.

\begin{figure}
\centerline{\includegraphics[width=0.45\textwidth]{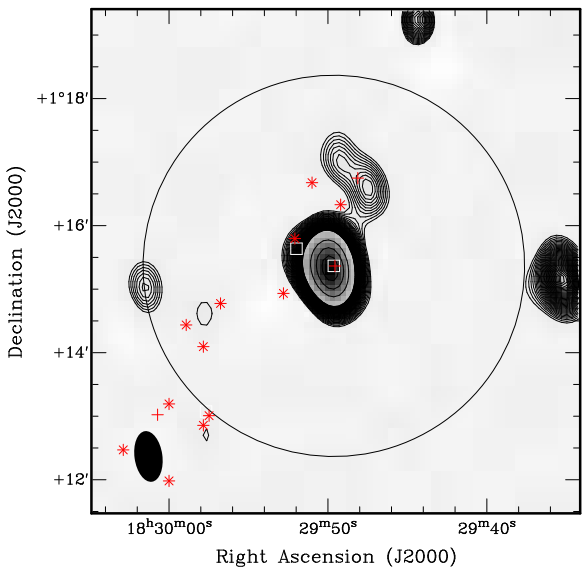}}
\caption{DCE08-210. Greyscale and contours are from the AMI-LA at 16\,GHz, with contours at 5,6,7,8\,$\sigma$ etc where $\sigma=19\,\mu$Jy\,beam$^{-1}$. Crosses (`+') mark the positions of Class~0 protostellar objects, stars (`$\ast$') mark the positions of Class~I objects, see Table~\ref{tab:serpa}. The position of candidate embedded objects from DCE08 are shown as squares. The AMI-LA primary beam FWHM is shown as a solid circle and the synthesized beam as a filled ellipse in the bottom left corner. \label{fig:210}}
\end{figure}

DCE08-211 (Serpens~SMM~10) lies to the north-east of DCE08-210, see Fig.~\ref{fig:210}. At such close separation the 16\,GHz data is still heavily dominated by the strong emission from DCE08-210 and it is not possible to attribute the eastern extension of this emission conclusively to DCE08-211 rather than the source VLA-8, which is positioned intermediate to DCE08-210 and DCE08-211.

To the north of DCE08-210 a spur of radio emission is observed at 16\,GHz. There are two distinct peaks within this spur which we attribute to HMH07-135 (Serpens~SMM~9; VLA-5) and HMH07-139. HMH07-135 was identified as Class~I by Kaas et~al (2004) and later as Class~0 by Enoch et~al. (2009). HMH07-139 (K254) was also identified as Class~I by Kaas et~al. (2004). HMH07-146 (VLA-9), found slightly to the east of these objects, is associated with Serpens~SMM~5; this source is undetected at 1.8\,cm with the AMI-LA.

At the eastern edge of this field there is a 16\,GHz radio point source which is not identified as protostellar in HMH07 or Evans et~al. (2009). This object may be extra-galactic, however it is coincident with Serpens~SMM~8, which is also coincident with a peak in CO emission (Davis et~al. 1999) and could therefore be a possible radio protostar. 

To the west of the AMI field is the bright radio source NVSS~J182934+011504. There is some debate over the nature of this object. Although it is coincident with a 2MASS point source it has no \emph{Spitzer} counterpart. Eiroa et~al. (1993) suggest that it is in fact extra-galactic, and indeed the flat spectral index measured between 3.5 and 1.8\,cm may be misleading for this object as the 1.4\,GHz flux density is dependent on resolution, varying by a factor of 1.5 between the NVSS catalogue (Condon et~al. 1998) and the higher resolution observations of Snell \& Bally (1986). This could be evidence of variability, but is more likely to arise as a result of flux lost from extended structure at high resolution as the object does appear extended in the AMI-LA 16\,GHz map. From the AMI-LA and NVSS data, which have more comparable visibility coverage than the AMI-LA and Eiroa et~al. (1993) data, we find $\alpha_{1.4}^{16}=0.58\pm0.03$.

\subsubsection{DCE08-215}

\begin{figure*}
\centerline{\includegraphics[width=0.5\textwidth]{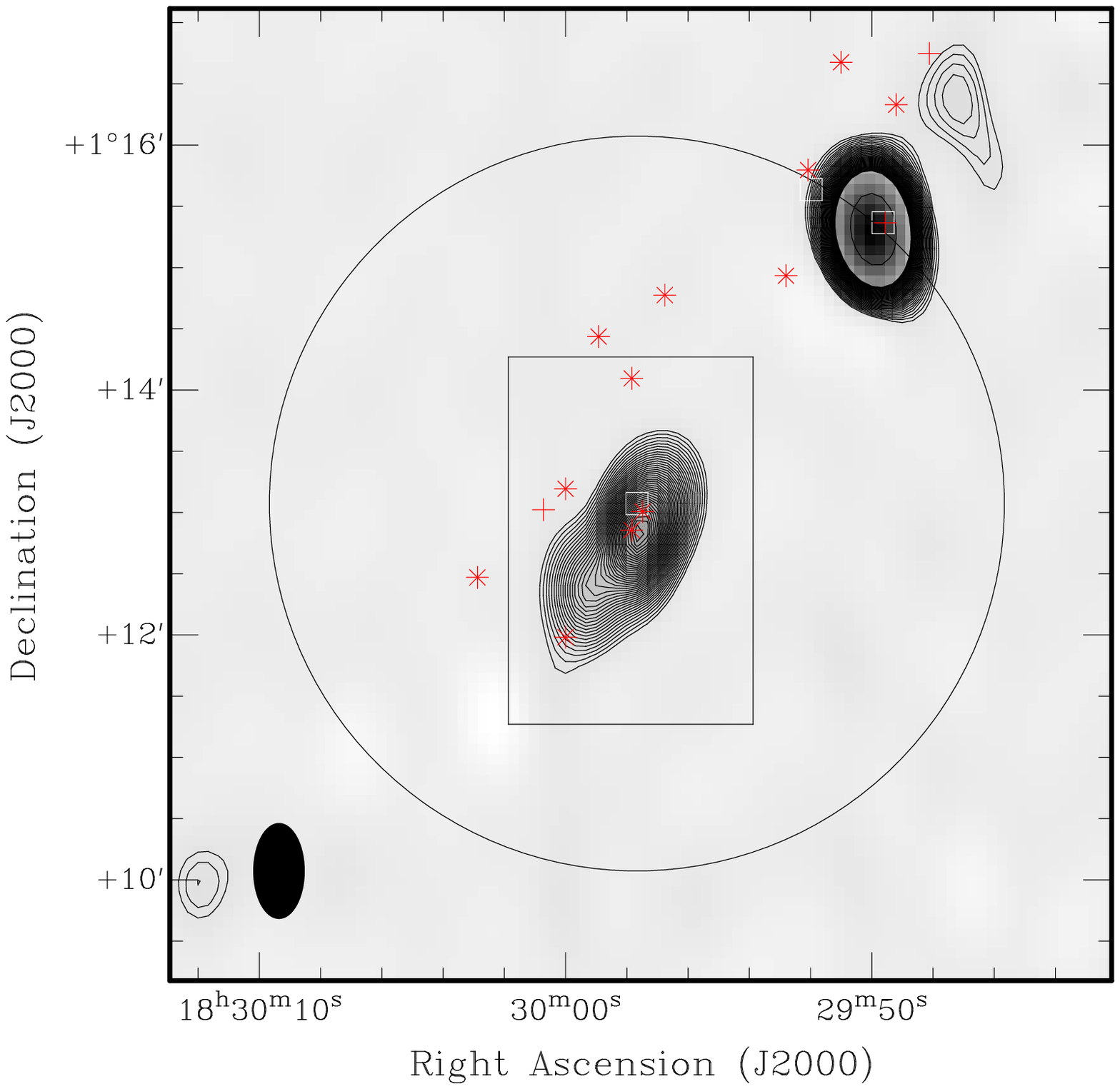}\qquad\includegraphics[width=0.44\textwidth]{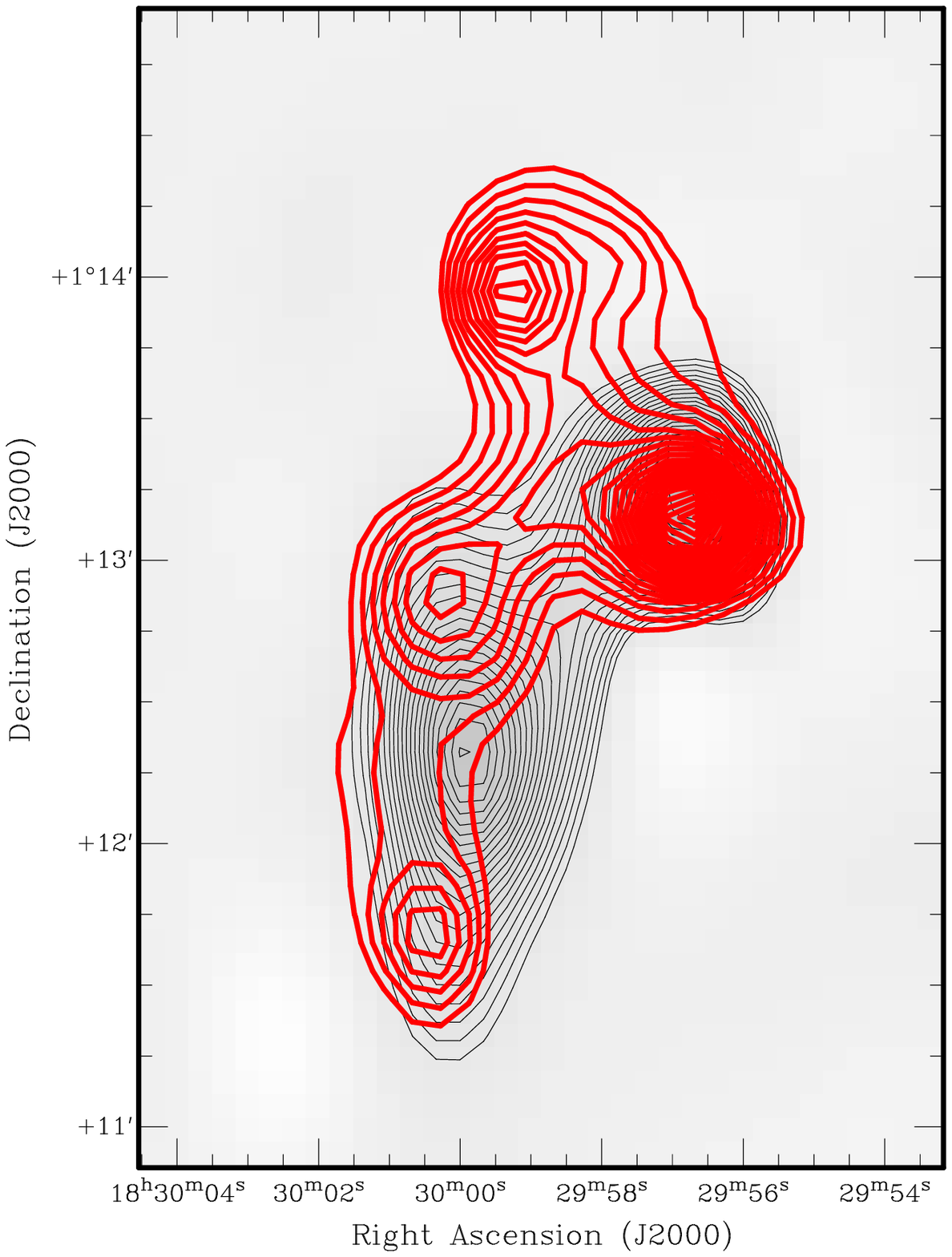}}

\centerline{(a) \hspace{0.5\textwidth} (b) }
\caption{(a) DCE08-215 field; (b) DCE08-215 field with point source subtraction. Greyscale and contours are from the AMI-LA at 16\,GHz, with contours at 5,6,7,8\,$\sigma$ etc where $\sigma=22\,\mu$Jy\,beam$^{-1}$.  Crosses (`+') mark the positions of Class~0 protostellar objects, stars (`$\ast$') mark the positions of Class~I objects, see Table~\ref{tab:serpa}. The position of candidate embedded objects from DCE08 are shown as squares. The AMI-LA primary beam FWHM is shown as a solid circle and the synthesized beam as a filled ellipse in the bottom left corner. The area shown in (b) is marked as a rectangle in (a). Thick contours in (b) are Scuba 850\,$\mu$m data from Di~Francesco et~al. (2008), shown in increments of 2~per~cent from a level of 15~per~cent.  \label{fig:215}}
\end{figure*}

The region around DCE08-215 contains several radio sources and protostellar sources, see Fig.~\ref{fig:215}(a). We observe a large kidney shaped extended source coincident with DCE08-215. This source is elongated north-south; Eiroa et~al. (1993) detect a number of resolved sources at higher resolution at 8.35\,GHz within this area and it is likely that the 16\,GHz source includes multiple contributions. The peak of the 16\,GHz radio emission is coincident with VLA-17 (Eiroa et~al. 1993) and we can assume that this source provides the dominant contribution to the emission. In order to examine the area in more detail we subtract a point source at this position directly from the visibility data with a flux density consistent with the 16\,GHz peak value. After removing this source the remaining emission has the morphology shown in Fig.~\ref{fig:215}(b). A distinct point source is seen to the north-west of the pointing centre, coincident with Serpens~SMM~4 (VLA-13); to the south-east an extended region of emission is present, coinciding with the sub-mm emission joining Serpens~SMM~2 and SMM~11. No 16\,GHz radio emission is seen towards Serpens~SMM~3. 

We associate the northern point source with K308, which was identified by Kaas et~al. (2004) as a Class~I object. It is not included in HMH07 due to the lack of a detection at 24\,$\mu$m. Conversely, Winston et~al. (2007) identify this object as Class~II, however here we retain the Class~I designation due to the presence of a sub-mm core and we identify this source with Ser-emb~22 (Enoch et~al. 2009). The extended emission to the south of DCE08-215 is more difficult to divide conclusively. We tentatively divide this emission between the two Class~I sources K331 (Kaas et~al. 2004) associated with VLA-19 (Eiroa et~al. 1993) and HMH07-198, and K330 (Kaas et~al. 2004) associated with HMH07-197.

\subsection{Cluster B Region}

\subsubsection{Existing radio observations of Serpens Cluster B}

Observations of the Serpens Cluster B region were made with the VLA at 3.5\,cm by Djupvik et~al. (2006). Of nine radio detections we associate four with protostellar objects from Harvey et~al. (2007). As with Cluster A the majority (three) of these radio sources correspond to Class 0/I objects; the remaining detection is associated with HMH07-59, which is identified as Class II/III (Harvey et~al. 2007; Harvey \& Dunham 2009). 

\subsubsection{DCE08-208}

The object DCE08-208 was identified as an external galaxy candidate by Evans et~al. (2009; [EDJ2009]-577) with a bolometric luminosity of 0.024\,L$_{\odot}$, but is considered a Group 1 candidate for an embedded protostellar source by DCE08. Harvey et~al. (2007a) who initially identify this object as HMH07-75, divide it further into three components A, B and C (Harvey et~al. 2007b). Of these components they make preliminary classifications based on bolometric temperature of Class I, I and 0, respectively. Component C has a 70\,$\mu$m counterpart, whereas components A and B do not, and it has been suggested that, although component A is point-like, component B may simply be extended emission from the outflow of component C. The combined bolometric luminosity of the three sources is heavily dominated by the Class 0 component, C, which has a re-derived value of 1.6\,L$_{\odot}$ (Harvey \& Dunham 2009). 

\begin{figure*}
\centerline{\includegraphics[width=0.5\textwidth]{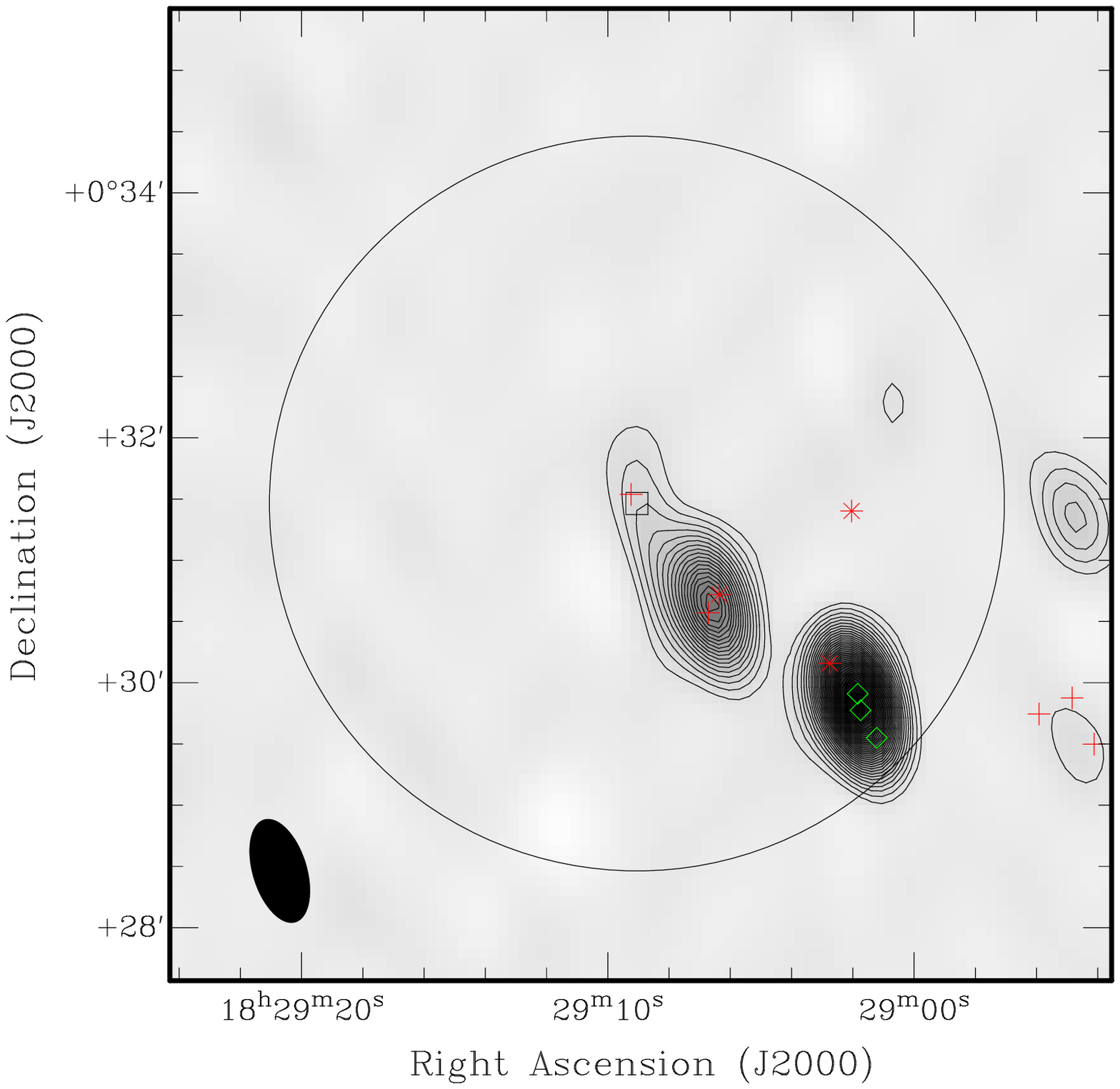}\qquad\includegraphics[width=0.5\textwidth]{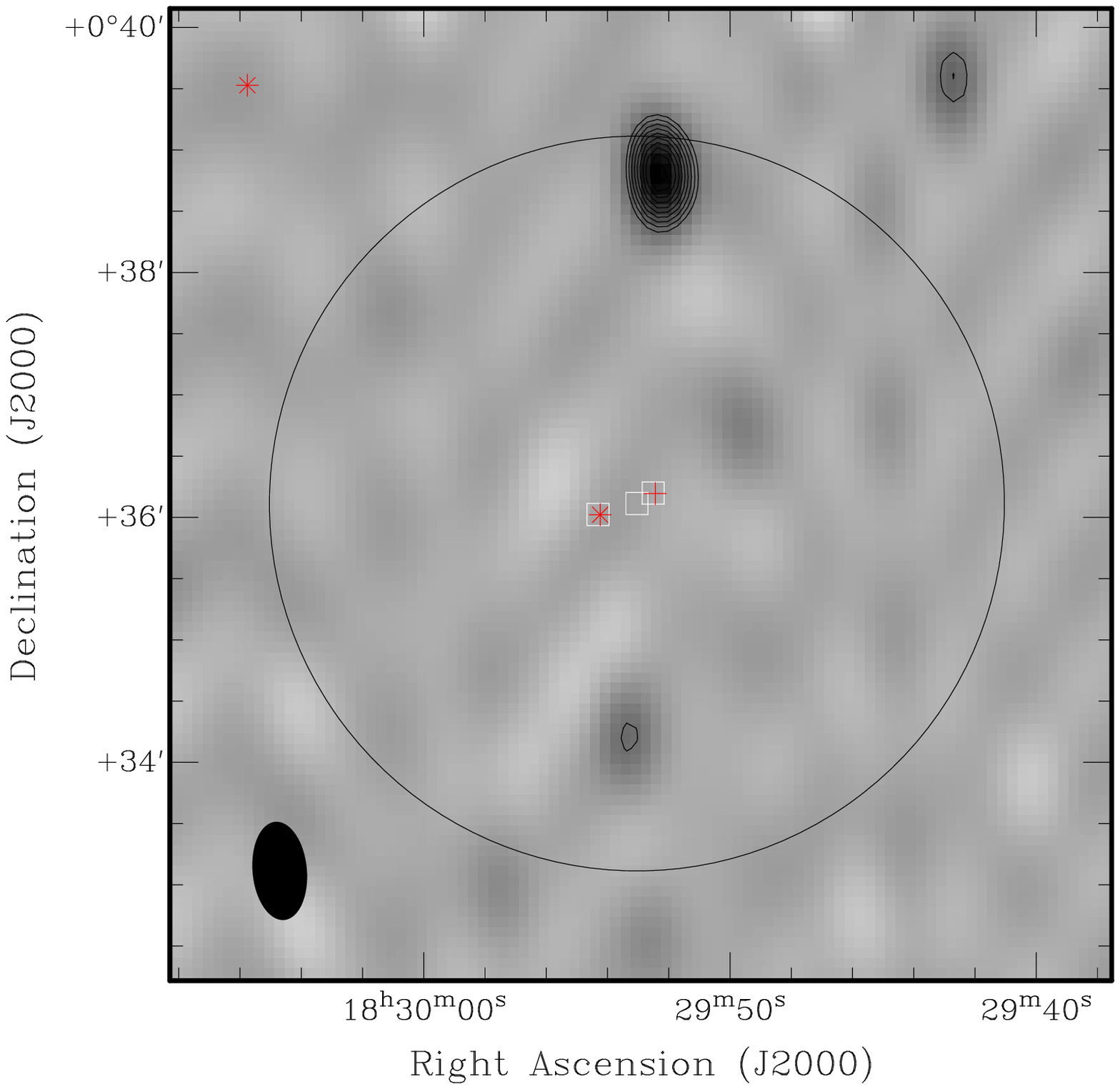}}

\centerline{(a) \hspace{0.5\textwidth} (b) }
\caption{Serpens Cluster B fields. (a) DCE08-208 field; (b) DCE08-213 field (no protostellar sources detected). Greyscale and contours are from the AMI-LA at 16\,GHz, with contours at 5,6,7,8\,$\sigma$ etc where $\sigma=16\,\mu$Jy\,beam$^{-1}$ for (a) and $\sigma=19\,\mu$Jy\,beam$^{-1}$ for (b).  Crosses (`+') mark the positions of Class~0 protostellar objects, stars (`$\ast$') mark the positions of Class~I objects, and diamonds (`$\diamond$') mark the positions of Class~II sources, see Table~\ref{tab:clb-obs}. The position of candidate embedded objects from DCE08 are shown as squares. The AMI-LA primary beam FWHM is shown as a solid circle and the synthesized beam as a filled ellipse in the bottom left corner. \label{fig:clusterb}}
\end{figure*}

Radio emission from DCE08-208 is detected at 16\,GHz as an extension to a brighter adjacent source, see Fig~\ref{fig:clusterb}(a). It is likely that this emission is dominated by HMH07-75C, and appears to be extended towards the north of this source. There is no corresponding radio source detected at 3.5\,cm (Djupvik et~al. 2006). Assuming that DCE08-208 was not significantly removed from the phase centre of the 3.5\,cm VLA observations this would give the radio emission a spectral index of $\alpha \geq 1.13$.

Immediately to the south-west of DCE08-208 is a bright radio source coincident with objects HMH07-67 \& -68. We associate this source with the Class 0 object HMH07-68 (Harvey \& Dunham 2009), based the on proximity of the radio peak to the \emph{Spitzer} position, however it may also have a contribution from the closely removed Class I YSO HMH07-67. HMH07-68 was also detected at 3.5\,cm by Djupvik et~al. (2006) and using their flux density we measure a spectral index for this object of $\alpha=2.07\pm0.40$. Such a steep rising spectral index is typical of optically thick free--free emission, such as is seen towards compact {\sc Hii} regions. These types of massive {\sc Hii} regions are not typical for Class~0 protostellar objects and it is therefore more likely that the significant increase in flux between 3.5\,cm and 1.8\,cm is due to the presence of either extended structure around this object which has been lost in the higher resolution VLA observations, or alternatively variability in the radio emission from HMH07-068.

Further to the south-west there is a third radio source, which is centred on the two sources HMH07-58 and -59. HMH07-59 was also detected at 3.5\,cm by Djupvik et~al. (2006) and it was suggested that the emission from this more evolved Class~II/III protostellar object could be due to gyro-synchrotron emission. If we assume that the flux density seen at 16\,GHz is entirely due to HMH07-59 we find that this object has a slightly rising spectrum with $\alpha=0.13\pm0.04$. Again such a spectrum could imply free--free as the dominant emission mechanism, however it is also not inconsistent with the previously proposed emission mechanism of gyrosynchrotron and this is discussed further in Section~\ref{sec:spindex}. The elongated nature of the detection suggests that it may also have contributions from HMH07-54 \& -56, which are again identified as Class~II/III objects. Fitting the 16\,GHz emission as the sum of two point sources, the first centred on HMH07-59 and the second on HMH07-56, we can constrain this additional emission to be $\leq 0.22$\,mJy. The fact that this second component was not detected at 3.5\,cm implies that it also possesses a rising spectrum. 

\subsubsection{DCE08-213}

There are two Class~0 objects close to the pointing centre of the DCE08-213 AMI-LA field. Neither of these is detected, and indeed no protostellar sources are detected at 16\,GHz in this field with flux densities $>96\,\mu$Jy. Two radio point sources are detected away from the pointing centre. The radio source to the north is identified with NVSS~J182952+003857 and has $S_{1.4\,{\rm{GHz}}}=2.8\pm0.6$\,mJy. Combining these data with the AMI-LA 16\,GHz flux density shows that this is a steep spectrum extra-galactic radio source with $\alpha=-0.58\pm0.07$. The second source has a flux density approximately half that of the northern source and, assuming a similar spectral index, this would place it below the detection limit in the NVSS survey at 1.4\,GHz. With no known protostellar object in the vicinity of this source we identify it as a non-thermal extra-galactic radio source.

\subsection{Serpens Filament}

\subsubsection{DCE08-202}

\begin{figure}
\centerline{\includegraphics[angle=0,width=0.5\textwidth]{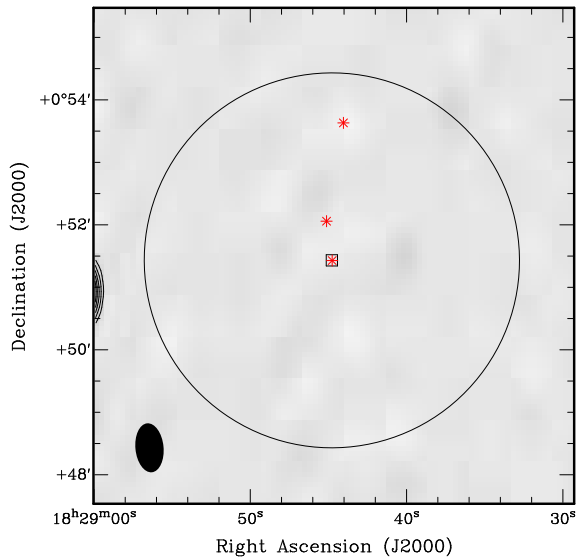}}
\caption{DCE08-202 (no protostellar sources detected). Greyscale and contours are from the AMI-LA at 16\,GHz, with contours at 5,6,7,8\,$\sigma$ etc where $\sigma=20\,\mu$Jy\,beam$^{-1}$ .  Stars (`$\ast$') mark the positions of Class~I objects, see Table~\ref{tab:filament}. The position of the candidate embedded object from DCE08 is shown as a square. No Class~0 objects are present in this field. The AMI-LA primary beam FWHM is shown as a solid circle and the synthesized beam as a filled ellipse in the bottom left corner \label{fig:202}}
\end{figure}

The DCE08-202 field in the Serpens filament is completely devoid of radio sources at 16\,GHz, see Fig.~\ref{fig:202}. The nearest source is almost 4\,arcmin to the east of the pointing field and is associated with NVSS~J182901+005117. Flux densities measured this far outside the AMI-LA primary beam are imprecise due to uncertainties in the primary beam model outside the FWHM but this source appears to have a flat spectrum $S_{1.4\,{\rm{GHz}}}=3.8\pm0.5$\,mJy, $S_{16\,{\rm{GHz}}}=3.5\pm0.4$\,mJy assuming 10\% calibration uncertainty at 16\,GHz gives $\alpha=-0.03\pm0.01$. 

The three protostellar objects in this field are all Class~I and are uniformly not detected. In the case of HMH07-26 this is surprising as its bolometric luminosity would suggest that it should be detectable according to the correlation between radio and bolometric luminosity measured in Paper II. However, it is immediately noticeable from Table~\ref{tab:filament} that although these objects vary significantly in bolometric luminosity they all have similarly low envelope mass estimates. In the case of HMH07-26, which when judged on its bolometric luminosity would have a 16\,GHz flux density of almost 300\,$\mu$Jy, has only a predicted 16\,GHz flux density of 38\,$\mu$Jy when judged on its envelope mass.

\subsection{Expected contamination by extragalactic radio sources}

At 16\,GHz we expect a certain number of extragalactic radio sources to be seen within each of our fields. Following Papers I \& II, to quantify this number we use the 15\,GHz source counts model from de Zotti et~al. (2005) scaled to the 10C survey source counts (Davies et~al. 2010).  The average rms noise from our datasets is $\simeq19\,\mu$Jy\,beam$^{-1}$ and from this model we predict that we should see 0.07 sources arcmin$^{-2}$, or $\simeq 2$ radio sources within a 6\,arcmin FWHM primary beam above a 5\,$\sigma$ flux density of 95\,$\mu$Jy. Within the Serpens core region we would therefore expect to detect $\approx 4\pm2$ extragalactic radio sources, and in the Serpens/G3-G6 region a further $\approx 9\pm3$ sources. Making the assumption that all sources which cannot be identified with a previously known protostellar object are extragalactic, as are sources with non-thermal spectra, we find 3 radio sources in the Serpens Cluster A fields, consistent with our prediction; in the remaining fields we find 3 further sources, which is lower than predicted, but consistent at 2\,$\sigma$. The low number of extra-galactic sources detected in these fields compared to the model may represent an over-estimation from the 15\,GHz source counts, which are extrapolated from a completeness limit of 0.5\,mJy - high compared with our detection limit of 95\,$\mu$Jy. One break in the observed source counts is already known at mJy flux density levels (Davies et~al. 2010) and it is quite possible that a further break exists below 0.5\,mJy.

\subsection{Expected radio flux density from thermal dust emission}
\label{sec:dust}

At 16\,GHz there is expected to be a small contribution to the radio flux density of protostars due to the long wavelength tail of the thermal dust emission from the envelopes around these young objects, and this contribution may be estimated using sub-mm data. However, the complexity and multiplicity of objects in Serpens makes accurate flux density calculation from sub-mm maps difficult; sources are crowded and several sources with near neighbours may be associated with the same sub-mm core. For the purposes of this work we therefore constrain the contribution of thermal dust emission to the measured 1.8\,cm flux densities of our detected sources by extrapolating modified greybody spectra for each object directly from 1.1\,mm flux densities. As the flux densities themselves are not listed by Enoch et~al. (2009a) we recover them from the envelope masses in that work  for Class~0 and Class~I objects in Serpens, following
\begin{equation}
S_{1.1\,{\rm{mm}}} = \frac{M_{\rm{env}}B_{1.1\,{\rm{mm}}}(T_{\rm{d}})\kappa_{1.1\,{\rm{mm}}}}{D^2},
\end{equation}
where $D=260$\,pc is the distance to Serpens, $\kappa_{1.1\,{\rm{mm}}}=0.0114$\,cm$^{2}$g$^{-1}$ is the opacity and $B_{1.1\,{\rm{mm}}}(T_{\rm{d}})$ is the Planck function at a dust temperature of $T_{\rm{d}}=15$\,K. The values of these parameters are chosen to match those used by Enoch et~al. (2009a) to derive envelope mass estimates from flux densities. The opacity is taken from the dust model of Ossenkopf \& Henning (1994) for grains which have coagulated over $10^5$ years and which have thin ice mantles (OH5; indicating Column [5] of Table 1 in Ossenkopf \& Henning 1994).

The source HMH07-139 (AMI-4) has no counterpart in the Bolocam catalogue of Enoch et~al. (2009a), however it is coincident with a distinct sub-mm peak in the higher resolution 1.3\,mm MAMBO map of Kaas et~al. (2004) where it is identified as a Class~I object. Following the method of Enoch et~al. (2009) we extract a sub-mm flux density for HMH07-139 by integrating the flux density within a $30''$ aperture, centred on the \emph{Spitzer} position, from the publically available 850\,$\mu$m data of Di~Francesco et~al. (2008). From these data we measure a flux density, $S_{850\,\mu{\rm{m}}}=1.9$\,Jy, and consequently an envelope mass of $M_{\rm{env}}=1.16$\,M$_{\odot}$, see Table~\ref{tab:serpa}.

Consequently, only one of the Class~0/I radio detections lacks a sub-mm flux density and envelope mass estimate, HMH07-186. This source is not identified with a distinct sub-mm peak, but instead is located between several adjacent compact cores. Without evidence for a peak at the position of this source we do not attempt to fit a flux density and envelope mass.

From the 1.1\,mm flux densities we estimate the thermal dust contribution at 16\,GHz by extrapolating a modified greybody model of the form $S_{\nu} = \nu^{\beta}B_{\nu}(T_{\rm{d}})$ from the 1.1\,mm measurement. To remain consistent with the envelope mass estimates we again use a dust temperature of $T_{\rm{d}}=15$\,K and an index of $\beta_{\rm{OH5}}=1.85$. Using an alternative dust temperature of 10\,K would increase the flux densities by approximately 30\%. The predictions themselves have uncertainties of 20-30\% from the errors on the 1.1\,mm flux densities alone. We note also that, as indicated in Ossenkopf \& Henning (1994), it is unlikely that $\beta$ will remain constant from 1.1\,mm to 1.8\,cm. However in the absence of additional data it is not currently possible to constrain this model further.

In what follows we use the measured flux density at 1.8\,cm with the predicted greybody contribution \emph{subtracted} to calculate the radio luminosity. This is to ensure that the values used are representative only of the radio emission and do not include contributions from the thermal dust tail which varies greatly between sources and which might therefore influence any conclusions being drawn from the correlations examined in the later stages of this paper. In general the corrections made to the radio flux densities for this sample are small, in the range $1-5$\%. Notable exceptions are AMI-3 and AMI-7 with corrections of 20 and 10~per~cent, respectively. We assume that any thermal dust contribution at 3.5\,cm is negligible and make no correction for this quantity in the flux densities taken from the literature.

\section{Correlations}
\label{sec:corr}

\begin{figure*}
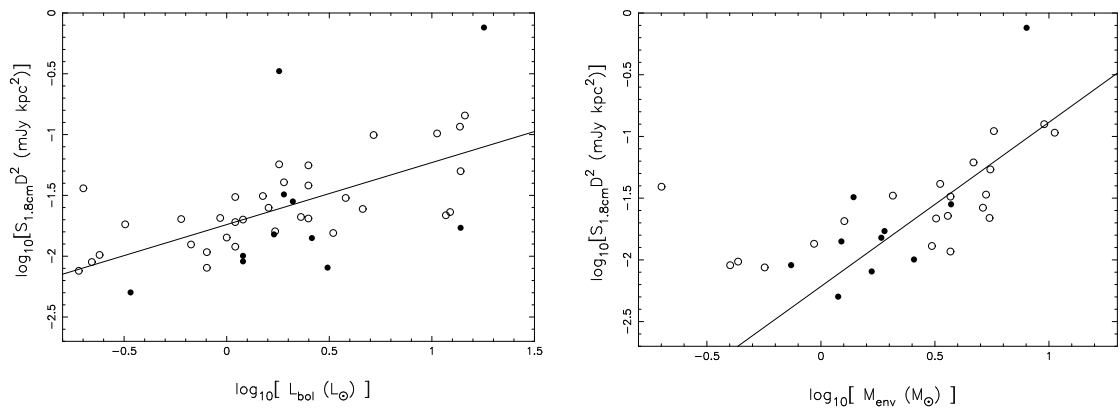

\centerline{\includegraphics[angle=-90,width=0.4\textwidth]{./eiroa_lbol_corr.ps}\qquad\includegraphics[angle=-90,width=0.4\textwidth]{./eiroa_menv_corr.ps}}
\caption{(a) Correlation of bolometric luminosity with 1.8\,cm radio luminosity; (b) correlation of envelope mass with 1.8\,cm radio luminosity. Unfilled circles show data from Papers I \& II in (a) and from Paper~II in (b). Filled circles show 3.5\,cm data from Eiroa et~al. and Djupvik et~al. 3.5\,cm radio flux densities have been scaled by a factor of 1.49, see text for details; masses from SCUBA data (H07; Paper~II) have been scaled by a factor of 0.33 to match those derived from Bolocam (E09), see text for details. Best fitting correlations from Paper~II are shown as solid lines. \label{fig:3.5corr}}
\end{figure*}

\begin{figure*}
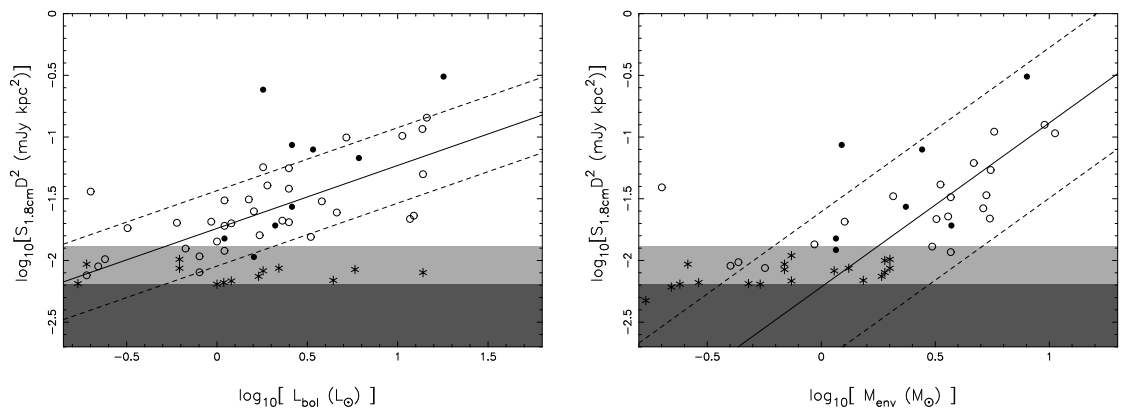

\centerline{\includegraphics[angle=-90,width=0.4\textwidth]{./serpens_lbol_corr.ps}\qquad \includegraphics[angle=-90,width=0.4\textwidth]{./serpens_menv_corr.ps}}
\caption{(a) Correlation of bolometric luminosity with 1.8\,cm radio luminosity; (b) correlation of envelope mass with 1.8\,cm radio luminosity, where masses from SCUBA data (H07; Paper~II) have been scaled by a factor of 0.33 to match those derived from Bolocam (E09), see text for details. Data points from Papers~I \&~II are shown as open circles; data points from this work are shown as filled circles for detected sources and stars (`$\ast$') at the position at the 5\,$\sigma$ upper limit for non-detections. Best fitting correlations from Paper~II are shown as solid lines, with the $1\,\sigma$ bounds on these fits indicated by dashed lines. Radio luminosities below the detection limit at the phase centre are shaded in dark grey, and the region shaded in light grey indicates the range of radio luminosities which lie between the detection limit at the phase centre and the detection limit at the half power point of the primary beam. \label{fig:1.8corr}}
\end{figure*}

We consider both the 3.5\,cm and 1.8\,cm data for these fields when forming correlations with the IR and sub-mm characteristics of the detected objects. In doing this we make two adjustments to the measured data. Firstly, for the 3.5\,cm flux densities we apply a scaling factor of $1.49=(3.5/1.8)^{0.6}$, in order to compare them directly with the 1.8\,cm flux densities. A spectral index of $\alpha=0.6$ corresponds to partially optically thick free--free emission from both a spherically symmetric density distribution (Panagia \& Felli 1975) and a collimated density outflow distribution (Reynolds 1986) and is generally assumed for the radio emission from protostellar objects. Secondly we scale the envelope mass values from Hatchell et~al. (2007) used in Paper~II by a factor of 0.33, in order to compare them directly with the mass values for objects in Serpens which are taken from the Bolocam catalogue of Enoch et~al. (2009). This factor is the average ratio of envelope masses for the sources from the Perseus region studied in Paper II when measured with Bolocam (Enoch et~al. 2009) and with SCUBA (Hatchell et~al. 2007). This factor accounts for differences in the assumed dust temperature, $T_{\rm{d}}$, the opacity, $\kappa_{\nu}$, after correcting for the assumed distance to Perseus, between these works when calculating mass values. We note that the limiting mass for the Enoch et~al. (2009) Bolocam catalogue is $M_{\rm{env}}\simeq 0.1$\,M$_{\odot}$, significantly below the lower limit indicated in Figs.~\ref{fig:3.5corr}~\&~\ref{fig:1.8corr}.

It can be seen, Fig.~\ref{fig:3.5corr}, that the 3.5\,cm data are consistent with the correlations measured in Papers~I \& II.
There are two notable outliers in Fig.~\ref{fig:3.5corr}, DCE08-210 and HMH07-59. Bolometric luminosity estimates for DCE08-210 (Serpens~SMM~1) vary; here we adopt a bolometric luminosity for this source of $L_{\rm{bol}} = 17.3$\,L$_{\odot}$ (Enoch et~al. 2009b). This value is higher than that determined solely from broadband sub-mm flux densities (Enoch et~al. 2009a, see Table~\ref{tab:serpa}) as it is calculated using the radiative transfer model of Dullemond \& Dominik (2004) and takes into account inclination effects. 

DCE08-210 is also detected in the NVSS survey at 1.4\,GHz (Condon et~al. 1998) with a flux density $S_{1.4\,{\rm{GHz}}}=4.5\pm0.5$\,mJy implying that it has a rising spectrum from 21\,cm to 1.8\,cm. The flux density measured at 3.5\,cm is in excess of both the 21\,cm and 1.8\,cm values, preventing a single power law fit. This may imply that DCE08-210 was ``flaring'' at the epoch of the 3.5\,cm VLA observations, which could explain the deviation of this datum from the general correlation. HMH07-59 is identified in Harvey \& Dunham (2009) as a Class~II/III source and consequently is the only object of its type included in this sample. It is not certain therefore that such an object should follow the general trend observed for Class~0 and Class~I objects.

\section{Discussion}
\label{sec:disc}

\subsection{Detections and non-detections}
\label{sec:det}

We show both the detected and undetected sources in Figs.~\ref{fig:1.8corr}(a) \& (b). In these plots data from Papers I and II are shown as unfilled circles, detections from this work are shown as filled circles, and non-detections from this work are shown as stars at the position of their 5\,$\sigma$ detection limits as listed in Tables~\ref{tab:serpa},~ \ref{tab:clb-obs} \& \ref{tab:filament}. The plot area below the average 5\,$\sigma$ detection limit at the field centres for the full sample of $95\,\mu$Jy is shaded in dark grey. The area corresponding to the detection limit between the field centres and the primary beam FWHM is shaded in light grey. Using the distribution of the data from Papers I \& II, 1\,$\sigma$ bounds are shown as dashed lines for the measured correlations, which are shown as solid lines. 

As has already been remarked upon, in Fig.~\ref{fig:1.8corr}(a) there are two detected sources which lie significantly above the bounds on the measured correlation. These are DCE08-210 (Serpens~SMM~1) and HMH07-59. Both of these objects also have discrepant 3.5\,cm flux densities, see Fig.~\ref{fig:3.5corr}(a). HMH07-59 is a Class~II/III source and, as the only source of this kind in the data, it is uncertain whether HMH07-59 should follow the general trend derived for Class~0/I objects. Since the radio emission mechanism for Class~II/III objects is unknown, and considered to be different from that of Class~0/I it seems possible that this object could deviate from the correlation. The 3.5\,cm flux density for DCE08-210 has already been suggested to represent a `flare' in the emission of this object, a possibility that is supported by the increased removal of this source from the correlation at 3.5\,cm compared with 1.8\,cm. The bolometric luminosity used here is $L_{\rm{bol}}=17.3$\,L$_{\odot}$, however significantly higher luminosities have been reported for this object which may explain the slight offset of this object from the trend.


In Fig.~\ref{fig:1.8corr}(a), there are a number of undetected sources which lie below the detection threshold, but outside the bounds on the general trend, in the region $L_{\rm{bol}}>2.0$\,L$_{\odot}$. There are four undetected sources in this region; three of these are located in multiple fields and in this case the limit corresponding to the field where the source is closest to the phase centre is plotted. Noticeably, however, this population does not appear correspondingly discrepant when considered in terms of their envelope mass, Fig.~\ref{fig:1.8corr}(b). A possible reason for this is the high ratio of bolometric luminosity to envelope mass measured for these objects. Excluding these four objects the mean ratio, $r = L_{\rm{bol}}/M_{\rm{env}}$, for this sample is $r = 1.13$. HMH07-182 and -176, which have the highest luminosities of those sources that are undetected, have $r=7.26$ and $r=8.41$, respectively. The remaining two objects in this region, HMH07-137 and -208, have $r=2.87$ and $r=1.67$, respectively.

Neglecting uncertainties in the assumed distance to Serpens, which are commented upon later, the errors on radio luminosity in these correlations are dominated by the conservative absolute calibration error of the AMI-LA of 5~per~cent. Uncertainties on the envelope mass have contributions from the assumed opacity index and temperature, as well as a 20-30~per~cent uncertainty from the sub-mm flux density, whereas errors on the bolometric luminosity, as stated in Enoch et~al. (2009) are approximately 10~per~cent.

\begin{figure}
\centerline{\includegraphics[angle=-90,width=0.4\textwidth]{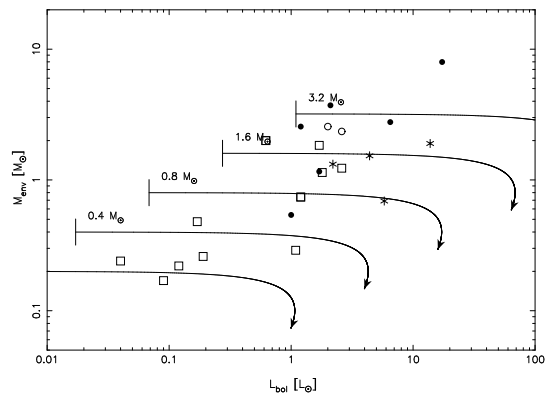}}
\caption{$M_{\rm{env}}$ vs. $L_{\rm{bol}}$ diagram for Serpens. Class~0 protostars are shown as filled circles and Class~I as open squares. Unclassified sources are shown as open circles. The four radio non-detections that are discrepant from the $L_{\rm{rad}}-L_{\rm{bol}}$ trend, but not the $L_{\rm{rad}}-M_{\rm{env}}$ trend are shown as stars. Evolutionary tracks are shown for time dependent accretion, $\dot{M}_{\rm{acc}} \propto M_{\rm{env}}(t)$, as described in the text, for initial mass reservoirs of $0.2-3.2$\,M$_{\odot}$, as indicated on the figure. \label{fig:evol}}
\end{figure}

The relationship of $L_{\rm{bol}}$ and $M_{\rm{env}}$ is often used to describe the evolution of embedded YSOs. The $L_{\rm{bol}}-M_{\rm{env}}$ diagram for this sample is illustrated in Fig.~\ref{fig:evol}. We make a simple illustration of the evolutionary behaviour of the YSOs in this diagram under the assumption that the accretion rate is proportional to the envelope mass, $\dot{M}_{\rm{acc}} \propto M_{\rm{env}}(t)$ (Bontemps et~al. 1996; Motte \& Andr{\'e} 2001). In this scenario collapse is initiated within a finite reservoir of material with mass $M^0_{\rm{env}}$, with accretion proceeding at a rate proportional to the mass of the residual envelope as a function of time, where $\dot{M}_{\rm{acc}} = M_{\rm{env}}/\tau$ and $\tau$ is a characteristic timescale taken in this instance to be $10^5$\,yrs. Folowing Bontemps et~al. (1996), the envelope mass is assumed to be exponentially decreasing and consequently the accretion rate has the functional form $\dot{M}_{\rm{acc}} = (M_{\rm{env}}^0/\tau) \exp{-t/\tau}$, and the stellar mass: $M_{\ast} = (M_{\rm{env}}^0)(1-\exp{-t/\tau})$.

Under a simple standard stellar model (e.g. Adams et~al. 1987) we may assume that during the infall stage the bolometric luminosity is heavily dominated by accretion, $L_{\rm{bol}} = L_{\rm{acc}} = GM_{\ast}\dot{M}_{\rm{acc}}/R_{\ast}$. We adopt a value of $R_{\ast} = 3$\,R$_{\odot}$ (Stahler 1988), but note that for a more sophisticated model the stellar radius will in fact be a function of both the accretion rate and the stellar mass, as will the stellar luminosity, which will also contribute an additive term to the bolometric luminosity such that $L_{\rm{bol}} = L_{\rm{acc}} + L_{\ast}$. However in this simple scenario where we limit ourselves to the earlier stages of evolution we consider this term to be negligible. Models which are required to provide information past the characteristic time-scale will require these additional terms (see e.g. Motte \& Andr{\'e} 2001).

Compared to the evolutionary diagrams for the Taurus and $\rho$-Ophiucus molecular clouds (Motte \& Andr{\'e} 2001) the protostellar objects in Serpens seem to have evolved from initial mass reservoirs which are a factor of $\approx 2$ larger. This apparent difference might be explained by the the fact that the entire envelope will not be completely accreted onto the central star, but that a proportion of it will instead be dispersed by the act of outflows (Ladd et~al. 1998). The four objects which are discrepant with the $L_{\rm{rad}}-L_{\rm{bol}}$ trend, but not the $L_{\rm{rad}}-M_{\rm{env}}$, have values of $r$ which place them in a similar region of the evolutionary diagram. Under our simple scenario they appear to be objects which have evolved from a relatively large initial mass reservoir and are approaching their characteristic timescale, where $\approx 50\%$ of their initial mass will have been accreted.

The detection statistics for all objects included in Tables~\ref{tab:serpa},~\ref{tab:clb-obs} and \ref{tab:filament} which lie within the FWHM of the AMI-LA primary beam and have an identified protostellar counterpart are summarized in Table~\ref{tab:detstat}. The small sky area, and consequently the small number of objects of each type present overall, means that no strong statistical conclusions can be drawn from this sample with the exception perhaps of the low Class~I detection rate for the extended sample. However, we note that the detection rates for Class~0 and Class~I objects in the extended sample are similar to those found for Perseus (Paper II) of 72 and 27\%, respectively.
\begin{table}
\begin{center}
\caption{Summary of detection statistics for Tables~\ref{tab:serpa},~\ref{tab:clb-obs} \& ~\ref{tab:filament}. The original sample includes only objects from DCE08, listed in Table~\ref{tab:sample}. The extended sample includes all sources from E09, inclusive of the original sample. \label{tab:detstat}}
\begin{tabular}{cccc}
\hline\hline
Class & Present & Detected & \% \\
\hline
\multicolumn{4}{l}{\emph{Original sample}:}\\
VeLLO & 0 & 0 & 0/100 \\
0 & 3 & 2 & 67 \\
I & 4 & 1 & 25 \\
Group 1 & 1 & 1 & 100\\
Group 2 & 4 & 1 & 25\\
Group 3 & 3 & 0 & 0\\
\multicolumn{4}{l}{\emph{Extended sample}:}\\
0 & 6  & 4 & 67 \\
I & 17 & 3 & 18 \\
\hline
\end{tabular}
\end{center}
\end{table}

\subsection{Outflow momentum fluxes}

\begin{figure}
\centerline{\includegraphics[angle=-90,width=0.4\textwidth]{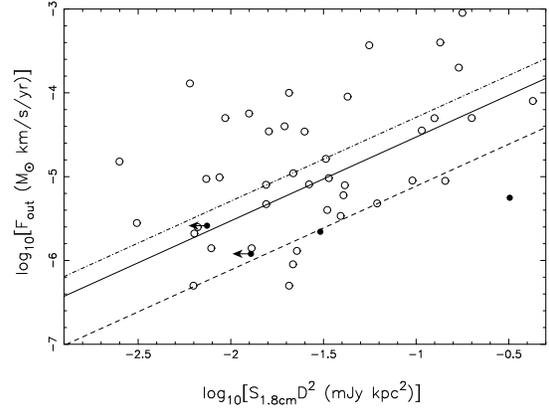}}
\caption{Correlation of radio luminosity with outflow momentum flux. Data from Papers~I \&~II are shown as unfilled circles, data from this work are shown as filled circles. The predicted relationship between momentum flux and radio luminosity (Curiel et~al. 1989) for a 100\% efficiency and a temperature of 3000\,K (dot-dash line), $10^4$\,K (solid line) and $10^5$\,K is indicated. \label{fig:foutcorr}}
\end{figure}

Graves et~al. (2010) calculate outflow momentum values for objects associated with four of the sub-mm cores in Serpens. Assuming a uniform dynamical age for each of $3\times10^4$\,years (Davis et~al. 1999) we can estimate the momentum flux for these sources as $F_{\rm{out}} = P/\tau$. The highly confused nature of this region precludes the use of more sophisticated momentum flux estimation methods. We use the maximum measured momentum from the blue and red outflow for each object, correcting the values in Table~3 of Graves et~al. (2010) by a factor of 2 to account for the full velocity range, as described in that paper; by a further factor of 10 to account for optical depth effects following Bontemps et~al. (1996) and Visser et~al. (2002); and by a small factor to account for the difference in distance assumed to Serpens. A comparison of these values with the overall correlation of radio luminosity to outflow momentum flux from Papers I \& II is shown in Fig.~\ref{fig:foutcorr}. As already described in Papers I \& II, following the model of Curiel et~al. (1989) a significant proportion of objects exhibit radio luminosities which are inconsistent with free-free emission arising solely as a result of shock ionisation in their molecular outflows. The four objects measured here (SMM~1, 4, 3/6 and 8) also appear to have radio emission inconsistent with this model, most markedly in the case of Serpens SMM~1. In Fig.~\ref{fig:foutcorr} we show the limit set by the shock ionisation model of Curiel et~al. (1989) on radio luminosity for a canonical electron temperature of $10^4$\,K, typical of free--free emission, as a solid line. Also indicated is the same model for a temperature of 3000\,K, the electron temperature indicated by $Herschel$ observations of shocks (van~Kempen et~al. 2010), and a temperature of $10^5$\,K. It is evident that even for very high temperature electron gas the shock ionisation model cannot account for the radio emission from a significant proportion of these objects.

However, the basis of a time dependent accretion rate, such as that outlined in the simple exponential model of \S~\ref{sec:det}, is founded on the assumption that there is a relationship between outflow force and accretion rate. This assumption arises from the hypothesis that the energy required to provide the outflow is released through accretion onto the central protostar (Bontemps et~al. 1996). If this is the case, and the observed correlation between radio luminosity, $S_{\nu}D^2$, and envelope mass shown in Fig.~\ref{fig:1.8corr} is a true one, then there should be a corresponding relationship between outflow momentum and radio luminosity. Therefore it seems likely that the lack of a strong correlation in Fig.~\ref{fig:foutcorr} is due to the inconsistent methods used to derive momentum flux across the sample.

\subsection{Distance dependence}
\label{sec:dist}

In the derived correlations and physical characteristics we have used a distance for the Serpens molecular cloud of 260\,pc, consistent with the \emph{Spitzer} catalogue. However, recently Dzib et~al. (2010) have suggested that Serpens has a distance of 415\,pc instead, using parallax measurements with the VLBA. This difference in assumed distance will affect the luminosity, mass and outflow momentum flux values quoted here. Radio luminosities will be raised by a factor of approximately 2.5, as will bolometric luminosities and envelope mass values, which are proportional to $D^2$. Since each of these quantities is increased proportionately to the other no change will occur in the degree of correlation shown in Figs.~\ref{fig:3.5corr} \&~\ref{fig:1.8corr}. Outflow momentum flux is linearly proportional to distance and consequently, the outflow momentum flux values shown in Fig.~\ref{fig:foutcorr} will be increased by a factor of 1.6, however the radio luminosity values will also be increased by a factor of 2.5. Since the increase in radio luminosity is proportionately greater than that of momentum flux this acts to worsen the agreement with the model (Curiel et~al. 1989) further.

\subsection{Radio spectral indices}
\label{sec:spindex}

The spectral index of partially optically thick free-free emission from fractionally ionized stellar winds and jets has been shown to vary depending on morphology (Reynolds 1986). Shallower spectral indices ($\alpha=0.25$) are found for collimated outflows than in the spherical case ($0.6<\alpha<0.8$), and steeper spectra still are found for conical jets ($\alpha>0.9$). In observations such as those presented here it is expected that unresolved contributions will be present in the radio spectrum from the core itself as well as a confined jet close to this core which then becomes unconfined at larger distances.

The mean value of spectral index for those sources which are detected at both 3.5 and 1.8\,cm is $\bar{\alpha}_{3.5}^{1.8}=0.53\pm1.14$, consistent with the canonical value for a partially optically thick spherical or collimated stellar wind of $\alpha=0.6$. However the possibility of variable radio emission from these sources may limit the meaningfulness of this agreement. Certainly in the case of Serpens~SMM~1 (DCE08-210), although we observe a non-thermal spectrum from 3.5 to 1.8\,cm, previous observations at a similar epoch measured an optically thin free--free spectrum. Across the AMI-LA band we see a similar behaviour with $\alpha_{\rm{AMI}}=-0.15\pm0.23$, consistent with optically thin free--free, although the short frequency coverage of the AMI-LA makes the uncertainty on this measurement large.

We do not fit spectral indices for AMI-7 and AMI-8 in the DCE08-215 field as the sources cannot be separated without the subtraction of AMI-8 from the map. This subtraction in itself will affect the spectrum of AMI-7. For the three radio detections in the DCE08-208 field we fit spectra across the AMI band, however the low signal to noise for these objects, combined with the short frequency coverage of AMI causes the uncertainties on these measurements to be large and consequently we suggest that they should only be used as an indication of emission mechanism rather than for quantitative comparisons. 

HMH07-59 (AMI-10) has a falling spectral index across the AMI band, with $\alpha_{\rm{AMI}} = -1.35\pm0.34$, differing from the optically thin index from 3.5 to 1.8\,cm. There are a number of possible explanations for this difference, notably the 3.5\,cm detection is associated solely to HMH07-59 whereas, although we also associate AMI-10 with this source, the 1.8\,cm emission is extended and probably has contributions from multiple sources. The extension of the emission itself will cause a steepening of the spectrum across the AMI band, as even a small amount of flux loss over such a short frequency range can cause significantly steepening. However, HMH07-59 is identified as a Class II/III object and therefore it is possible that the emission we are detecting is not thermal bremsstrahlung but gyrosynchrotron. Gyrosynchrotron emission is characterised by a sharply peaked spectrum, where the frequency of the peak is strongly dependent on the magnitude of the magnetic field, $B$. Following Dulk, Melrose \& White (1979) the position of this peak can be determined by
\begin{eqnarray}
\footnotesize
\nonumber \left[\frac{\nu_{\rm{peak}}}{\rm{GHz}}\right]  \simeq  &2.4& \left[\left(\frac{n_{\rm{e}}}{10^{10}{\rm{cm}}^{-3}}\right) \left(\frac{d}{10^8{\rm{cm}}}\right) \right]^{0.1}\\
\nonumber &&\times (\sin{\theta})^{0.6} \left[\frac{T_{\rm{e}}}{10^8{\rm{K}}}\right]^{0.7} \left[\frac{B}{10^2{\rm{G}}}\right]^{0.9},
\end{eqnarray}
where is also a weak dependency on the electron density, $n_{\rm{e}}$, and the path length, $d$, as well as the electron temperature of the plasma $T_{\rm{e}}$ and the inclination angle between the wave-normal direction and the magnetic field, $\theta$. A flat spectral index from 3.5 to 1.8\,cm and a falling spectral index across the AMI band implies that the peak of such a gyro-synchrotron spectrum would be found in the region of $\nu\simeq12$\,GHz. The electron density in such objects is assumed to be in the range $10^3\leq n_{\rm{e}} \leq 10^9$\,cm$^{-3}$, and the path length, roughly equivalent to the infall radius, to be in the range $0.01 \leq d \leq 0.1$\,pc. Making the assumptions of $n_{\rm{e}}=10^5$\,cm$^{-3}$ and $d=0.1$\,pc, as well as assuming an electron temperature of $10^8$\,K and an inclination angle of $\theta=\pi/4$, this would imply a magnetic field strength of $B=270$\,G, with approximately one order of magnitude uncertainty from the possible variation in parameters, as well as a small variation in the exponents dependent on the exact optical depth (Dulk, Melrose \& White 1979).

Such a magnetic field is not unlikely in the vicinity of a Class II/III object. Typical magnetic field strengths of 10s of Gauss have been suggested at a distance of a few astronomical units for objects with circumstellar disks (Shu et~al. 2007), and much higher fields of a few kiloGauss have been measured from the surfaces of classical T-Tauri stars (Donati et~al. 2005; Johns-Krull, Valenti \& Koresko 1999; Johns-Krull 2007). More recently kiloGauss fields have also been measured from a Class I object (WL~17; Johns-Krull et~al. 2009). A hypothesis of gyrosynchrotron emission from this object would be well tested by full radio polarization measurements, as gyrosynchrotron emission is known to be highly circularly polarized. As the AMI-LA measures I+Q it is unable to provide any information on the Stokes~V component towards this object.

The remaining two sources, HMH07-68 (AMI-10) and HMH07-75 (AMI-11), both appear to have optically thin spectra with $\alpha_{\rm{AMI}} = -0.17\pm1.08$ and $\alpha_{\rm{AMI}}=-0.05\pm1.15$, respectively. In the case of HMH07-68 this may indicate that the optical depth of the ionized medium has reached unity between 3.5 and 1.8\,cm, typical of ultra-compact {\sc Hii}.

The conclusions drawn from comparisons of $\alpha_{3.5}^{1.8}$ and $\alpha_{\rm{AMI}}$ are questionable as the 3.5 and 1.8\,cm observations have been made at widely different epochs. The question of variability in the radio emission of protostars is still unanswered and any variation over the period intervening these measurements will make these comparisons invalid.

\subsection{Evidence for radio variability in DCE08-210 (Serpens~SMM~1)}

Serpens~SMM~1 was previously observed at 15\,GHz by Snell \& Bally (1986) who found a flux density of $S_{15}=10\pm3$\,mJy. Disregarding the large uncertainty in this measurement, it differs from the AMI-LA 16\,GHz flux density by a factor of $\approx 2$. The 5\,GHz flux density measured by Rodr{\' i}guez et~al. (1989) is $S_{5}=6.4\pm0.4$\,mJy, a factor of 1.5 lower than that of Snell \& Bally (1986). Additional evidence for variability may also be identified when comparing the flux densities from Snell \& Bally (1986) with that at 8.5\,GHz from Eiroa et~al. (1993). Although the spectral index, $\alpha_{5}^{15}=0.0\pm0.3$, measured by Snell \& Bally (1986) from observations made at a similar epoch implies optically thin free-free emission, the 8.5\,GHz flux density is discrepant from this model by $\approx5\,\sigma$. Similarly the 25.4\,GHz flux density of Ungerechts \& Gusten (1984) of $S_{25.4}=18\pm2$\,mJy is significantly removed from the predictions of this model, in the opposite sense, by a degree which cannot be accounted for by an enhanced thermal dust contribution which is $<1\,$mJy.

\begin{figure}
\centerline{\includegraphics[angle=-90, width=0.4\textwidth]{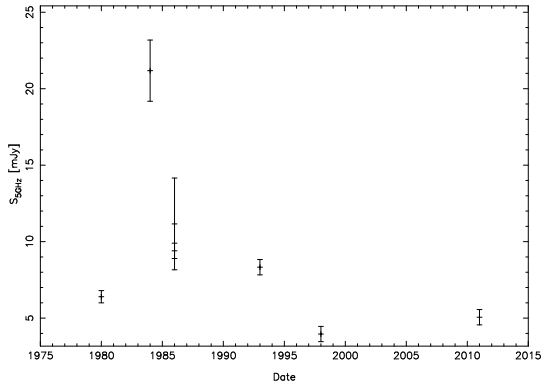}}
\caption{Radio flux density of Serpens~SMM~1 adjusted to 5\,GHz, see text for details, at different epochs. \label{fig:variable}}
\end{figure}

Evidence for variable accretion in DCE08-210 has previously been suggested by Enoch et~al. (2009b) who detect a high mass disk around this source, which should be unstable and therefore exhibit luminosity variations. The radio data for this object suggest that there was previously a radio ``flare" event, the additional emission from which has now faded. We show this possibility in Fig.~\ref{fig:variable}, where we have assumed that the spectrum of DCE08-210 is optically thin from 1.4 to 22\,GHz in a single epoch and have corrected the measured flux densities to a common frequency of 5\,GHz using a canonical spectral index for optically thin free--free emission of $\alpha=-0.1$.

\section{Conclusions}
\label{sec:conc}

The aim of this sample was to increase the data for Class~I objects and improve the mass to radio luminosity correlation measured in Paper~II. The low detection rate for Class~I objects in Perseus is however echoed in Serpens and only 18\% of Class~I objects present have been detected (3 sources). Although this is only a small increase in the available data these objects do not demonstrate the same evolutionary divide hinted at in the correlation of radio flux density with envelope mass seen in Paper~II. The objects detected in Serpens do however follow the general trends measured for radio luminosity with both bolometric luminosity and envelope mass, and the respective detection rates for Class~0 and Class~I objects are consistent with those found for Perseus in Paper~II. We also conclude that envelope mass provides a better radio detection indicator than bolometric luminosity with the deviations from the $L_{\rm{rad}}-M_{\rm{env}}$ correlation being notably fewer than that of the $L_{\rm{rad}}-L_{\rm{bol}}$ correlation.

\section{ACKNOWLEDGEMENTS}
We thank the staff of the Lord's Bridge Observatory for their
invaluable assistance in the commissioning and operation of the
Arcminute Microkelvin Imager. The AMI-LA is supported by Cambridge
University and the STFC. CRG, TS, TF, MO and MLD   
acknowledge the support of PPARC/STFC studentships. AS would like to acknowledge support from Science 
Foundation Ireland under grant 07/RFP/PHYF790.

\end{document}